\documentclass[twocolumn]{aastex61} 

\usepackage{hyperref,enumerate}
\input{colordvi}
\listofchanges

\hypersetup{
    unicode=false,                  
    pdftoolbar=true,                
    pdfmenubar=true,                
    pdffitwindow=true,              
    pdfstartview={FitH},            
    pdftitle={R443130},             
    pdfauthor={vplacco},            
    pdfsubject={Astronomy},         
    pdfcreator={dvipdf},            
    pdfproducer={dvipdf},           
    pdfkeywords={metal-poor stars}, 
    pdfnewwindow=true,              
    colorlinks=true,                
    linkcolor=red,                  
    citecolor=blue,                 
    filecolor=magenta,              
    urlcolor=cyan,                  
    breaklinks=true,
    linktocpage
}

\newcommand{\eps}[1]{\ensuremath{\log\epsilon\,(\mathrm{#1})}}
\newcommand{\xx}{{\tablenotemark{a}}}
\newcommand{\yy}{{\tablenotemark{b}}}
\newcommand{\zz}{{\tablenotemark{c}}}

\newcommand{{\rii}}{{\object{RAVE~J2038$-$0023}}}
\newcommand{\rII}{\object{RAVE~J203843.2$-$002333}}
\newcommand{\abund}[2]{\ensuremath{[\mathrm{#1}/\mathrm{#2}]}}
\newcommand{\cfe}{\abund{C}{Fe}}

\newcommand{\xfe}[1]{\abund{#1}{Fe}}
\newcommand{\metal}{\abund{Fe}{H}}

\newcommand{\teff}{\ensuremath{T_\mathrm{eff}}}
\newcommand{\logg}{\ensuremath{\log\,g}}

\shorttitle{A Highly R-process-Enhanced Star from the RAVE Survey}
\shortauthors{Placco et al.}

\begin{document}

\title{{\rII}: The First Highly R-process-Enhanced Star \\ Identified in the
RAVE Survey.\footnote{Based on observations gathered with the $6.5\,$m Magellan
Telescopes located at Las Campanas Observatory, Chile; Kitt Peak National
Observatory, National Optical Astronomy Observatory (NOAO Prop. ID: 14B-0231;
PI: Placco), which is operated by the Association of Universities for Research
in Astronomy (AURA) under cooperative agreement with the National Science
Foundation. The authors are honored to be permitted to conduct astronomical
research on Iolkam Du'ag (Kitt Peak), a mountain with particular significance
to the Tohono O'odham.}}

\author{Vinicius M.\ Placco}
\altaffiliation{Visiting astronomer, Kitt Peak National Observatory, National
Optical Astronomy Observatory, which is operated by the Association of
Universities for Research in Astronomy (AURA) under a cooperative agreement
with the National Science Foundation.} 
\affiliation{Department of Physics, University of Notre Dame, Notre Dame, IN 46556, USA}
\affiliation{JINA Center for the Evolution of the Elements, USA}

\author{Erika M.\ Holmbeck}
\affiliation{Department of Physics, University of Notre Dame, Notre Dame, IN 46556, USA}
\affiliation{JINA Center for the Evolution of the Elements, USA}

\author{Anna Frebel}
\affiliation{Department of Physics and Kavli Institute for Astrophysics and
Space Research, Massachusetts Institute of Technology, Cambridge, MA 02139, USA}
\affiliation{JINA Center for the Evolution of the Elements, USA}

\author{Timothy C.\ Beers}
\affiliation{Department of Physics, University of Notre Dame, Notre Dame, IN 46556, USA}
\affiliation{JINA Center for the Evolution of the Elements, USA}

\author{Rebecca A.\ Surman}
\affiliation{Department of Physics, University of Notre Dame, Notre Dame, IN 46556, USA}
\affiliation{JINA Center for the Evolution of the Elements, USA}

\author{Alexander P.\ Ji}
\affiliation{Department of Physics and Kavli Institute for Astrophysics and
Space Research, Massachusetts Institute of Technology, Cambridge, MA 02139, USA}
\affiliation{JINA Center for the Evolution of the Elements, USA}

\author{Rana Ezzeddine}
\affiliation{Department of Physics and Kavli Institute for Astrophysics and
Space Research, Massachusetts Institute of Technology, Cambridge, MA 02139, USA}
\affiliation{JINA Center for the Evolution of the Elements, USA}

\author{Sean D. Points}
\affiliation{Cerro Tololo Inter-American Observatory, Casilla 603, La Serena, Chile}

\author{Catherine C. Kaleida}
\affiliation{Space Telescope Science Institute, 3700 San Martin Drive, Baltimore, MD 21218, USA}

\author{Terese T. Hansen}
\affiliation{Observatories of the Carnegie Institution of Washington, 
813 Santa Barbara St., Pasadena, CA 91101, USA}

\author{Charli M. Sakari}
\affiliation{Department of Astronomy, University of Washington, Seattle, WA 98195-1580, USA}

\author{Andrew R. Casey}
\affiliation{Monash Centre for Astrophysics, School of Physics and Astronomy, 
Monash University, 19 Rainforest Walk, Vic 3800, Australia}

\correspondingauthor{Vinicius M.\ Placco}
\email{vplacco@nd.edu}

\begin{abstract}

We report the discovery of \rII, a bright ($V = 12.73$), very metal-poor
(\metal~=~$-$2.91), \emph{r}-process-enhanced (\abund{Eu}{Fe}~=~$+$1.64 and
\abund{Ba}{Eu}~=~$-$0.81) star selected from the RAVE survey.  This star was
identified as a metal-poor candidate based on its medium-resolution ($R \sim
1,600$) spectrum obtained with the KPNO/Mayall Telescope, and followed-up with
high-resolution ($R \sim 66,000$) spectroscopy with the Magellan/Clay
Telescope, allowing for the determination of elemental abundances for 24
neutron-capture elements, including thorium and uranium.  {\rii} is only the
fourth metal-poor star with a clearly measured U abundance.  The derived
chemical-abundance pattern exhibits good agreement with those of other known
highly \emph{r}-process-enhanced stars, and evidence in hand suggests that it
is not an actinide-boost star.  Age estimates were calculated using Th/X and U/X
abundance ratios, yielding a mean age of $13.0\pm 1.1$ Gyr.

\end{abstract}


\section{Introduction}
\label{intro}

Advances in observations and theory in the past few years are converging on
identifying the likely astrophysical site(s) of the rapid neutron-capture
process (\emph{r}-process), some sixty years after it was first suggested to
account for the production of roughly half of the heavy elements beyond iron
\citep{b2fh,cameron1957}.  The recent discovery of highly
\emph{r}-process-enhanced stars in the ultra-faint dwarf galaxy Reticulum~II
\citep{ji2016,roederer2016b} opens a new observational window on the origin of
the \emph{r}-process.  The observed enhancements point to enrichment by a rare
astrophysical event that copiously produces r-process elements. 
The presently favored site that fits these characteristics \citep[high
temperatures, densities, and flux of free neutrons on short
timescales;][]{b2fh} is the outflow from binary neutron star mergers
\citep[NSMs;][]{meyer1989,bauswein2013,rosswog2014}.  This environment has been
argued to be a possible source of the \emph{r}-process since the work of
\citet{lattimer1974}. 
If this hypothesis is correct, it would be possible to link all the
\emph{r}-process-enhanced stars observed to date (including those in the halo
field) to a common formation site and/or class of parent progenitors, which
would add important constraints to theoretical predictions for the chemical
evolution of the Galaxy and the Universe. Other possible sites of the
\emph{r}-process, including the so-called magneto-rotational supernovae
(MR-SNe), which address several concerns raised about NSMs as the single site
(e.g., \citealt{cescutti2015}; \citealt{tsujimoto2015}; \citealt{wehmeyer2015};
\citealt{beniamini2016}), are currently being explored \citep{nishimura2017}. 

Observations of stars in ultra-faint dwarf galaxies are challenging
due to their faint magnitudes ($g \gtrsim 17$). Because of that, (brighter)
field halo stars can provide more detailed information on the \emph{r}-process
element abundances, to help better constrain its origins.
The modern era of detailed exploration of this question opened with the
discovery of the highly \emph{r}-process-enhanced star CS~22892-052
\citep{sneden1994}, an extremely metal-poor star \citep[originally identified in
the HK Survey of Beers and collaborators;][]{beers1985,beers1992} with
\emph{r}-process elemental-abundance ratios exceeding ten times the solar
values. These stars are known as \emph{r}-II stars \citep[\xfe{Eu} $> +1.0$ and
\abund{Ba}{Eu} $< 0.0$;][]{beers2005}.  Other examples of such stars have been
identified over the past few decades, as the result of dedicated searches
\citep[e.g., HERES, the Hamburg/ESO R-process Enhanced Star survey,
see][]{christlieb2004,barklem2005} and other large high-resolution spectroscopic
studies of very metal-poor \citep[VMP; \metal\footnote{\abund{A}{B} =
$\log(N_A/{}N_B)_{\star} - \log(N_A/{}N_B) _{\odot}$, where $N$ is the number
density of atoms of a given element in the star ($\star$) and the Sun ($\odot$),
respectively.} $< -2.0$;][]{beers2005,frebel2015} and extremely metal-poor (EMP;
[Fe/H] $< -3.0$) stars in the Galactic halo
\citep[e.g.,][]{cayrel2004,roederer2014}, and now number on the order of 25
stars.

The remarkable agreement between the \emph{r}-process-element pattern observed
in \emph{r}-II stars and the Solar System suggests that either the
\emph{r}-process elements 
were well-mixed into the interstellar medium, or more likely, that the production of
\emph{r}-process elements resulted from the contribution by a unique
astrophysical site in the early Galaxy. 
Furthermore, suggestions that the
\emph{r}-process enhancement in stars could be the result of peculiarities in the
atmospheres of evolved stars or associated with mass-transfer binaries have been
disproven as a result of (i) The identification of \emph{r}-process-enhanced
stars in essentially all stages of stellar evolution \citep{roederer2014b} and
(ii) The binary frequency of such stars revealed by long-term radial-velocity
monitoring \citep[18 $\pm$ 6 \%;][]{hansen2015b}  being similar to the
frequency of other halo stars lacking this signature \citep[16 $\pm$ 4
\%;][]{carney2003}.

The identification of \emph{r}-II stars requires high-resolution spectroscopy.
Among the $\sim$25 \emph{r}-II stars with published analyses
\citep{saga2008,frebel2010b}, the abundances of \emph{both} thorium and uranium
could only be measured in three cases (CS~31082-001; \citealt{hill2002},
HE~1523-0901; \citealt{frebel2007}, and CS~29497-004; \citealt{hill2016}). The
star BD+17$^{\circ}$3248 \citep [\xfe{Eu} = $+0.9$;][]{cowan2002} is considered
by the authors to have a tentative U detection. A higher quality
spectrum of this star is needed to better constrain the U abundance.  The
abundances of radioactive isotopes of elements such as Th and U can also place
constraints on the age of the Universe, and be used to validate their early
production, within the first $\sim$0.5-1.5 Gyr following the Big Bang. Age
estimates are obtained by application of the nucleo-chronometry technique,
pioneered for metal-poor stars by \citet{butcher1987}, using theoretical
production ratios and abundance ratios of stable \emph{r}-process elements and
radioactive isotopes (e.g., $^{232}$Th, half-life $14.0$ Gyr, and $^{238}$U,
half-life $4.5$ Gyr). In the case that both U and Th are measured in the star,
the U/Th chronometer pair can be used \citep{cayrel2001, hill2002,hill2016}. 

\begin{deluxetable*}{lr|lrrrr}
\tablewidth{0pt}
\tabletypesize{\small}
\tablecaption{Observational Data \label{candlist}}
\tablehead{
\multicolumn{2}{c}{{\rii}} &&
Mayall &
Magellan 2014 &
Magellan 2016 &
RAVE}
\startdata
$\alpha$ (J2000) &    20:38:43.2 & Date          & 2014 09 15 & 2014 09 25 & 2016 04 16 &   \nodata \\
$\delta$ (J2000) & $-$00:23:33   & UT            &   02:19:52 &   04:17:03 &   08:46:14 &   \nodata \\
$V$ (mag)        &         12.73 & Exptime (s)   &        600 &        900 &      5,400 &   \nodata \\
$B-V$            &          0.99 & \emph{R}      &      2,000 &     38,000 &     66,000 &     8,000 \\
$g$ (mag)        &         13.32 & V$_{r}$(km/s) &   $-$332.9 &   $-$321.7 &   $-$321.6 &  $-$319.6 \\
$g-r$            &          0.87 & S/N (3860\AA) &         50 &         30 &        100 &   \nodata \\
$J$ (mag)        &         10.73 & S/N (4550\AA) &         80 &         90 &        220 &   \nodata \\ 
$J-K$            &          0.41 & S/N (7900\AA) &    \nodata &        150 &    \nodata &   \nodata \\ 
\enddata
\end{deluxetable*}

In this paper we report the discovery of the \emph{r}-II star {\rII} (hereafter
{\rii}; \metal$ = -2.91$), the fourth low-metallicity star where abundances of
both Th and U could be confidently measured. This star was originally selected as
a bright ($V = 12.7$) VMP candidate from the RAVE \citep[RAdial Velocity
Experiment;][]{steinmetz2006} fourth data release \citep[DR4;
][]{kordopatis2013}\footnote{A later data release, DR5 \citep{kunder2017},
published after the analysis presented in the present work, provides refined
parameter estimates.}, and medium-resolution
spectroscopy with the KPNO/Mayall telescope revealed that this target is indeed
a low-metallicity giant without carbon enhancement. Subsequent high-resolution
follow-up with the MIKE spectrograph on the Magellan/Clay Telescope confirmed
the presence of enhancements in \emph{r}-process elements, such as Ba, Eu, Th,
and U, which are reported here.

This paper is outlined as follows. Section~\ref{secobs} describes the target
selection for the medium-resolution spectroscopic investigation and the
high-resolution follow-up observations, followed by the determinations of the
stellar parameters in Section~\ref{secatm}. Section~\ref{secab} provides
details on the abundance determinations. Section~\ref{seccomp} discusses the
\emph{r}-process abundance pattern of {\rii} compared with other \emph{r}-II
stars, including those with previously detected U, and obtains age estimates
for {\rii} based on selected chronometry pairs. Our conclusions and a brief
discussion are provided in Section~\ref{final}.

\section{Target Selection and Observations}
\label{secobs}

{\rii}\ was selected as a metal-poor candidate star from RAVE DR4, part of a
sub-sample with $4500 < \teff < 5750$ and \metal\ $< -1.8$. These targets were
then followed up with medium-resolution spectroscopy on a variety of telescopes,
in order to validate their atmospheric parameters and obtain carbon abundance
estimates. High-resolution spectroscopic follow-up was then carried out for the
most interesting candidates. The full description of the target selection and
spectroscopic follow-up will be provided in a forthcoming paper. 

\subsection{Medium-Resolution Spectroscopy}

Medium-resolution spectroscopic follow-up was carried out with the
Mayall 4-m Telescope at Kitt Peak National Observatory. The observations were
obtained in semester 2014B, using the R-C spectrograph, with the KPC007 grating
(632~l~mm$^{\rm{-1}}$), the blue setting, a 1$\farcs$0 slit, and covering the
wavelength range [3500,6000]\,{\AA}. This combination yielded a resolving power
of $R\sim 1,600$, and signal-to-noise ratio S/N $\sim 80$ per pixel at 4550\,
{\AA}. The calibration frames included FeAr exposures (taken following the
science observation), quartz-lamp flat-fields, and bias frames. All reduction
tasks were performed using standard
IRAF\footnote{\href{http://iraf.noao.edu}{http://iraf.noao.edu}.} packages.
Table \ref{candlist} lists details of the observations from RAVE, and also the
medium- and high-resolution spectroscopic follow-ups. 

\begin{figure*}[!ht]
\epsscale{1.05}
\plotone{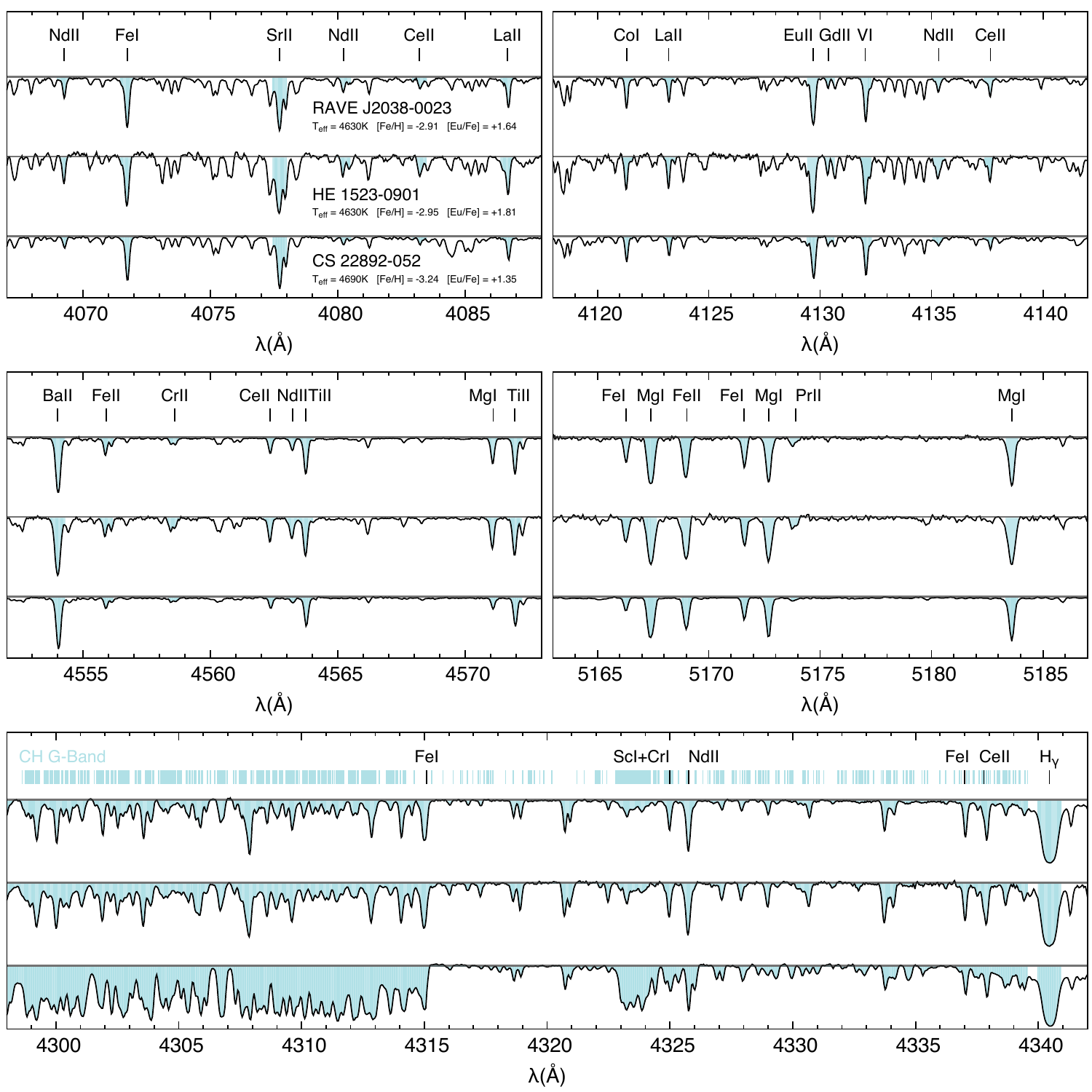}
\caption{Selected regions of the spectrum of \protect{\rii}, compared with the
\emph{r}-II stars HE~1523$-$0901 \citep{frebel2007} and CS~22892$-$052
\citep{roederer2014}. The upper and middle panels show regions where absorption
features of Fe, Mg, and neutron-capture elements are present. The lower panel
shows the CH $G$-band, used for carbon abundance determinations.
\explain{changed color of CH features on bottom panel}}
\label{highres}
\end{figure*}

\subsection{High-Resolution Spectroscopy}

High-resolution spectroscopic data  were obtained during the 2014B
and 2016A semesters, using the Magellan Inamori Kyocera Echelle
\citep[MIKE;][]{mike} spectrograph on the Magellan/Clay Telescope at Las
Campanas Observatory. For the 2014B run, the observing setup included a
$0\farcs7$ slit with $2\times2$ on-chip binning, yielding a resolving power of
$R\sim38,000$ (blue spectral range) and $R\sim30,000$ (red spectral range),
measured from the arc lamp spectral features. The
S/N at $4550\,${\AA} is $\sim$90. MIKE spectra have nearly full optical
wavelength coverage ([$\sim$3500,9000]\,{\AA}). For the 2016A run, the
observations were carried out using the $0\farcs35$ slit with $2\times2$
on-chip binning, yielding a resolving power of $R\sim66,000$. The S/N at
$3860\,${\AA} (close to the U spectral feature) is $\sim$100, and $\sim$220 at
$4550\,${\AA} (near a prominent \ion{Ba}{2} feature). 
The data were reduced using the data reduction pipeline developed for MIKE
spectra, described by
\citet{kelson2003}\footnote{\href{http://code.obs.carnegiescience.edu/python}
{http://code.obs.carnegiescience.edu/python}}.  Figure~\ref{highres} shows the
spectrum of {\rii}, compared with the \emph{r}-II stars HE~1523$-$0901
\citep[\teff=4630~K, \metal=$-2.95$, and \xfe{Eu}=$+1.81$;][]{frebel2007} and
CS~22892$-$052 \citep[\teff=4690~K, \metal=$-3.24$, and
\xfe{Eu}=$+1.35$;][]{roederer2014}, in regions where absorption features of
neutron-capture elements are present (upper and middle panels), as well as in
the region of the molecular CH $G$-band feature (lower panel). 

\section{Stellar Parameters}
\label{secatm}

\subsection{Medium-Resolution Spectrum}

The stellar atmospheric parameters (\teff, \logg, and [Fe/H]), and the carbon
abundance from the medium-resolution spectrum were obtained using the n-SSPP, a
modified version of the SEGUE Stellar Parameter Pipeline
\citep[SSPP;][]{lee2008a,lee2008b,lee2013}. The values for \teff, \logg, and
\metal, determined from photometry, line-indices, and matching with a
synthetic spectral library \citep[see][for further details]{beers2014}, were
used as first estimates for the high-resolution analysis. Results are listed in
Table \ref{obstable}.

\subsection{High-Resolution Spectra}

From the high-resolution MIKE spectrum, we determined the stellar parameters
spectroscopically (see details below), using the SMH software developed
by \citet{casey2014}.  Equivalent-width measurements were obtained by fitting
Gaussian profiles to the observed absorption lines within SMH. Table
\ref{eqw} lists the lines used in this work, their measured equivalent widths,
and the derived abundance from each line. We employed one-dimensional
plane-parallel model atmospheres with no overshooting \citep{castelli2004},
computed under the assumption of local thermodynamic equilibrium (LTE).

\begin{deluxetable}{llccc}[!ht]
\tablewidth{0pc}
\tabletypesize{\scriptsize}
\tablecaption{Derived Stellar Parameters for \protect\rii \label{obstable}}
\tablehead{
\colhead{              }&
\colhead{\teff{}(K)    }&
\colhead{\logg{}(cgs)  }&
\colhead{\metal{}      }&
\colhead{$\xi$(km/s)   }}
\startdata
RAVE (DR4) & 4315 (105)	& 0.82 (0.40) &	$-$2.17 (0.10) & \nodata	 \\
RAVE (DR5) & 4502 (51)	& 1.18 (0.21) &	$-$2.60 (0.14) & \nodata     \\
RAVE-on    & 4801 (82)	& 1.49 (0.15) &	$-$2.74 (0.07) & \nodata     \\
KPNO       & 4655 (150) & 0.85 (0.35) & $-$3.10 (0.20) & \nodata     \\
Magellan   & 4630 (100) & 1.20 (0.20) & $-$2.91 (0.10) & 2.15 (0.20) \\
\enddata
\end{deluxetable}

The effective temperature of {\rii} was determined by minimizing trends between
the abundances of 202 \ion{Fe}{1} lines and their excitation potentials, and
applying the temperature correction to the photometric scale suggested by
\citet{frebel2013}. The microturbulent velocity was determined by minimizing the
trend between the abundances of \ion{Fe}{1} lines and their reduced equivalent
widths. The surface gravity was determined from the balance of the two
ionization stages of iron, \ion{Fe}{1} and \ion{Fe}{2}. {\rii}\ also had its
stellar atmospheric parameters determined from the moderate-resolution
($R\sim8,000$) RAVE spectrum by \citet{kordopatis2013}. These values, together
with our determinations from the medium- and high-resolution spectra, are listed
in Table \ref{obstable}. For completeness, we also include the parameters from
the most recent release, RAVE DR5 \citep{kunder2017}, and also from the RAVE-on
catalog \citep{casey2016}.

There is very good agreement between the effective temperatures derived from the
medium- and high-resolution spectra used in this work; the RAVE DR5 value is
less than $\sim$150~K cooler. The surface gravities are all within 1$\sigma$,
and the high-resolution and RAVE DR5 $\log g$ values are nearly identical. The
surface gravity estimates from RAVE DR4 and KPNO (\logg~= 0.82 and 0.85
respectively) are expected to be similar, as both of these estimates come from
isochrone matching, while the high-resolution estimate (\logg~= 1.20) was
determined spectroscopically. The {\metal} estimate from RAVE DR4 appear
significantly higher than those reported from  RAVE DR5 and the medium- and
high-resolution results; the latter two of which are in good agreement with one
another. The RAVE-on metallicity is in better agreement with our high-resolution
estimate than either DR4 or DR5 from RAVE. However, despite the RAVE-on result
having a reduced chi-squared value of 0.63, this star was excluded from the
RAVE-on release because the RAVE pre-processing pipeline, SPARV, flagged \rii\
as being a star with much higher temperature (\teff$  > $10,000~K).

\begin{figure*}[!ht]
\centering
\begin{minipage}{0.32\textwidth}
\centering
\includegraphics[width=\textwidth]{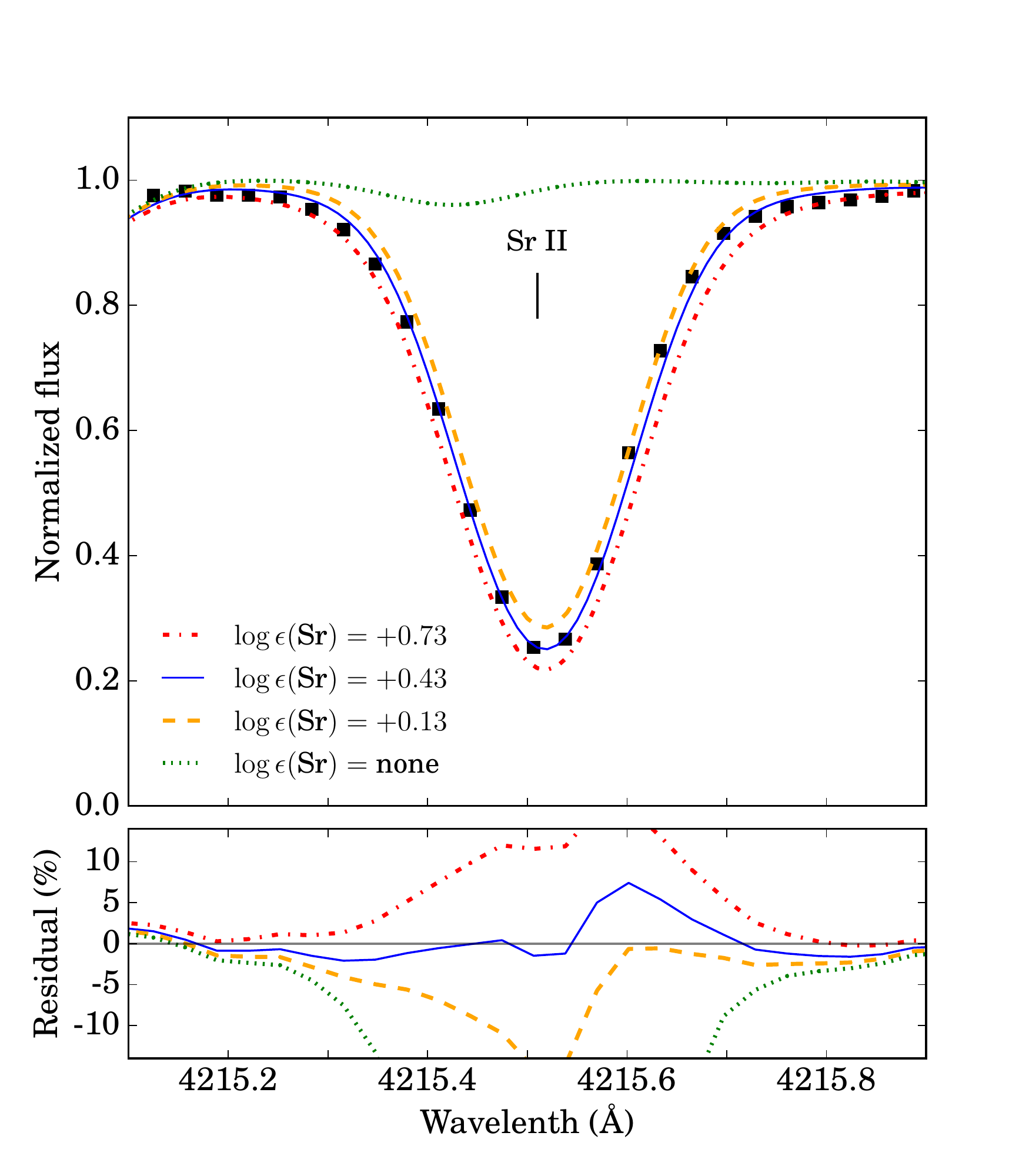}
\end{minipage}%
\begin{minipage}{0.32\textwidth}
\includegraphics[width=\textwidth]{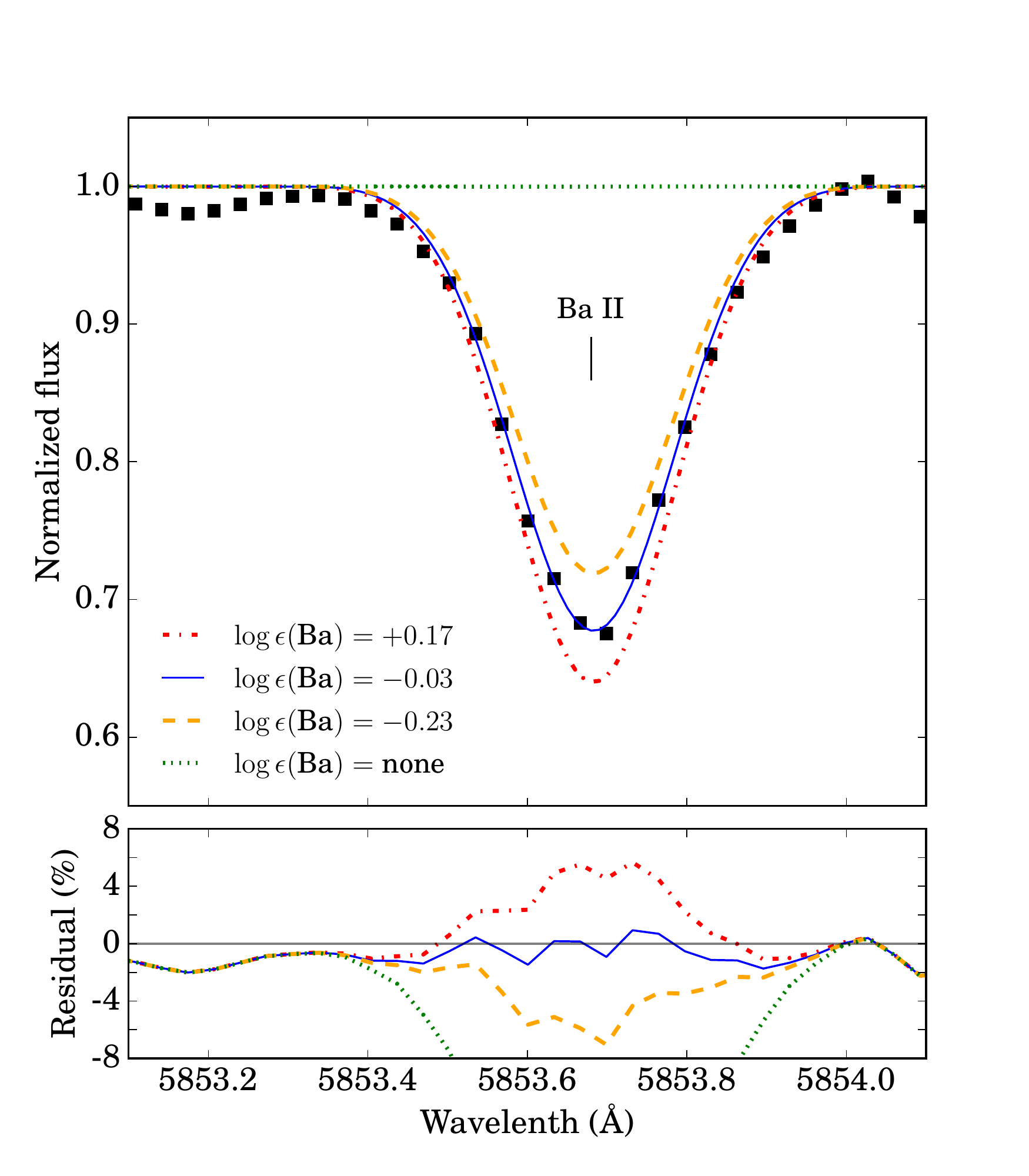}
\vfill
\end{minipage}
\begin{minipage}{0.32\textwidth}
\includegraphics[width=\textwidth]{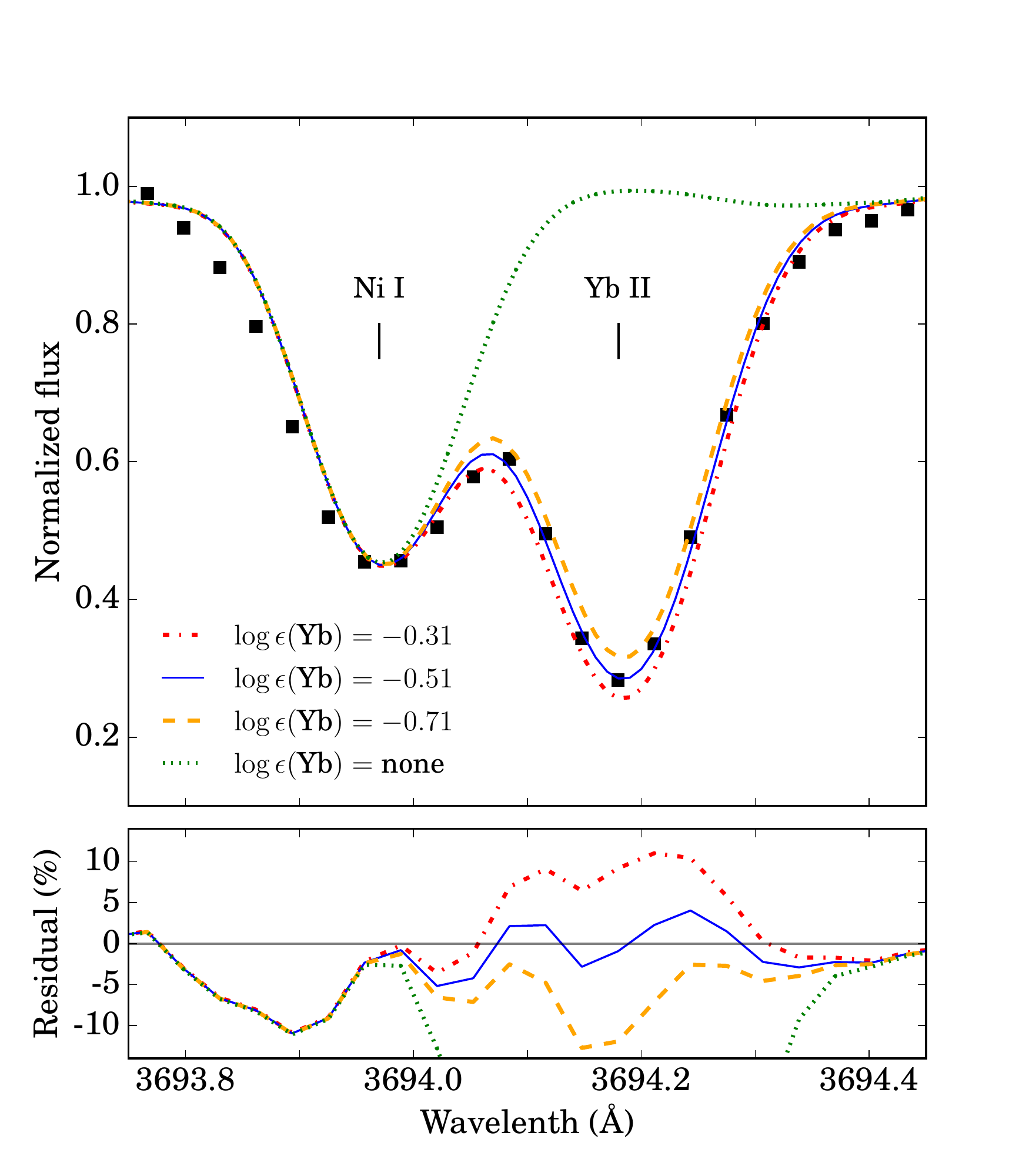}
\vfill
\end{minipage}
\caption{Observed spectra (squares) and synthesis (lines) for varying abundances
of \ion{Sr}{2} (left panel), \ion{Ba}{2} (middle panel), and \ion{Yb}{2} (right
panel).}
\label{fig:synfig}
\end{figure*}

\section{Chemical Abundances} \label{secab}

Chemical abundances for {\rii}\ were calculated by equivalent-width analysis and
spectral synthesis, using the MOOG code \citep[July 2014
version;][]{sneden1973}, which includes a proper treatment of scattering
\citep[see][for details]{sobeck2011}. The set of atmospheric parameters used for
the abundance analysis is the one derived from the Magellan spectra. Tables
\ref{eqw} and \ref{tab:ncap} list the derived abundances for individual lines
for light elements (C--Zn) and neutron-capture elements (Sr--U), respectively.
The excitation potentials and oscillator strengths for the lines employed
are taken from a variety of sources, including the compilations of
\citet{aoki2002}, \citet{barklem2005}, and \citet{roederer2012d}, as well as
from the VALD database \citep{vald} and the National Institute of Standards
and Technology Atomic Spectra Database \citep[NIST; ][]{nist}.
Elemental-abundance ratios, \xfe{X}, are calculated adopting solar
photospheric abundances from \citet{asplund2009}. The average abundances for
$39$ elements, derived from the Magellan/MIKE spectra, are listed in Table
\ref{abfinal}. The $\sigma$ values are the standard deviation and the
$\overline{\sigma}$ are the standard error of the mean. For elements where
$\overline{\sigma}$ is lower than 0.10, we adopt a fixed value of 0.10
\citep[see discussion in Section 4.6 of][]{placco2013}.

Uncertainties in the abundance determinations, as well as the systematic
uncertainties due to changes in the atmospheric parameters, were treated using
the same procedures described in \citet{placco2013,placco2015b}. Table \ref{sys}
shows the changes in the derived chemical abundances due to variations (within
the quoted uncertainties) in each atmospheric parameter. Also listed is the
total uncertainty, calculated from the quadratic sum of the individual
estimates. This calculation used only spectral features with abundances
determined by equivalent-width analysis. The variations are $+$100~K for \teff,
$+$0.2~dex for \logg, and $+$0.2 km\,s$^{-1}$ for $\xi$.

\subsection{C to Zn}

The carbon abundance for {\rii}\ was derived from the CH molecular feature at
$\lambda$4313 (\cfe\ $ = -0.44$).  Since this star is on the
upper red-giant branch, the measured carbon abundance does not reflect the
chemical composition of its natal gas cloud. 
Using the procedure described in \citet{placco2014}, we determined that the
expected carbon depletion due to CN processing for {\rii}\ is 0.67~dex.
Taking this into account, the corrected value for the carbon abundance is \cfe\
= $+0.23$.  Abundances of Na, Mg, Al, Si, Ca, Sc, Ti, V, Cr, Mn, Co, Ni, and Zn
were determined by equivalent-width analysis and spectral synthesis. Individual
line determinations are listed in Table \ref{eqw}, and final abundances are
provided in Table \ref{abfinal}.

\subsection{Neutron-Capture Elements}

The chemical abundances for the neutron-capture elements were determined via
spectral synthesis performed using MOOG. The results for individual lines are
given in Table \ref{tab:ncap}. Below we provide details on these measurements.
Note that the uncertainty on individual synthesis measurements is
typically set as $\pm$0.2~dex, as the measured abundance is well-bound between these
limits (see, e.g., Figure \ref{fig:synfig}).

\begin{figure*}[!ht]
\centering
\begin{minipage}{0.32\textwidth}
\centering
\includegraphics[width=\textwidth]{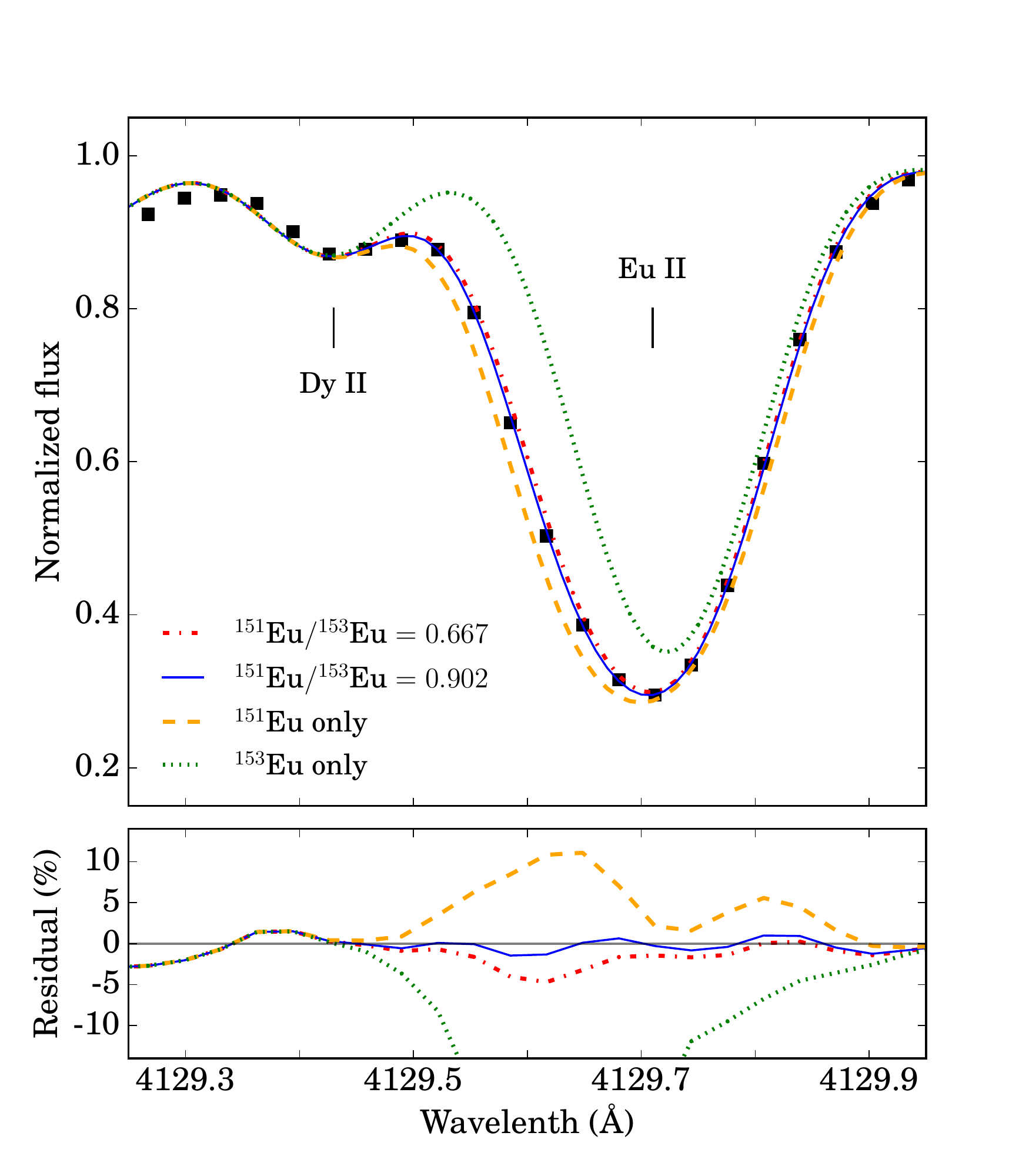}
\end{minipage}%
\begin{minipage}{0.32\textwidth}
\includegraphics[width=\textwidth]{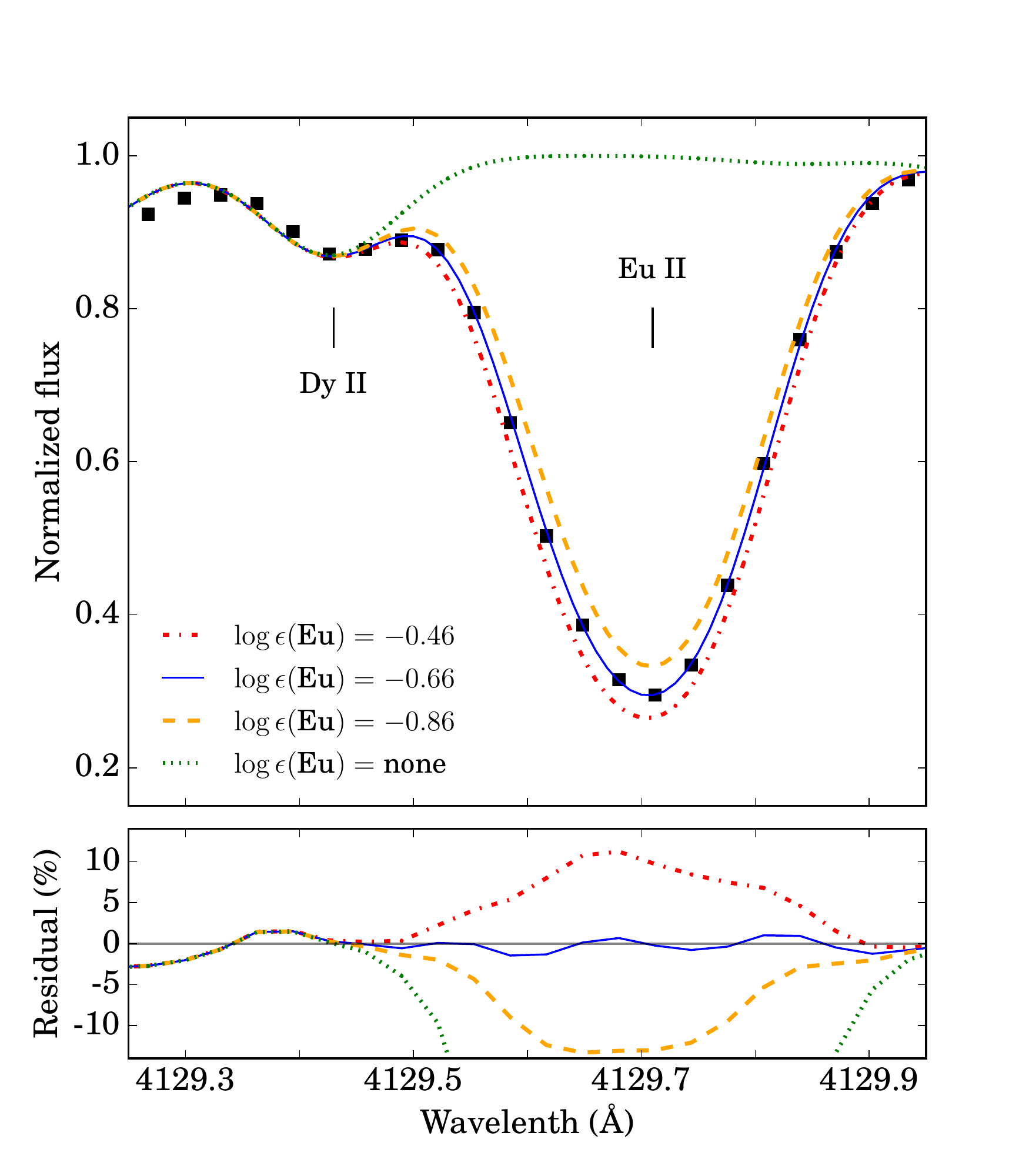}
\vfill
\end{minipage}
\begin{minipage}{0.32\textwidth}
\includegraphics[width=\textwidth]{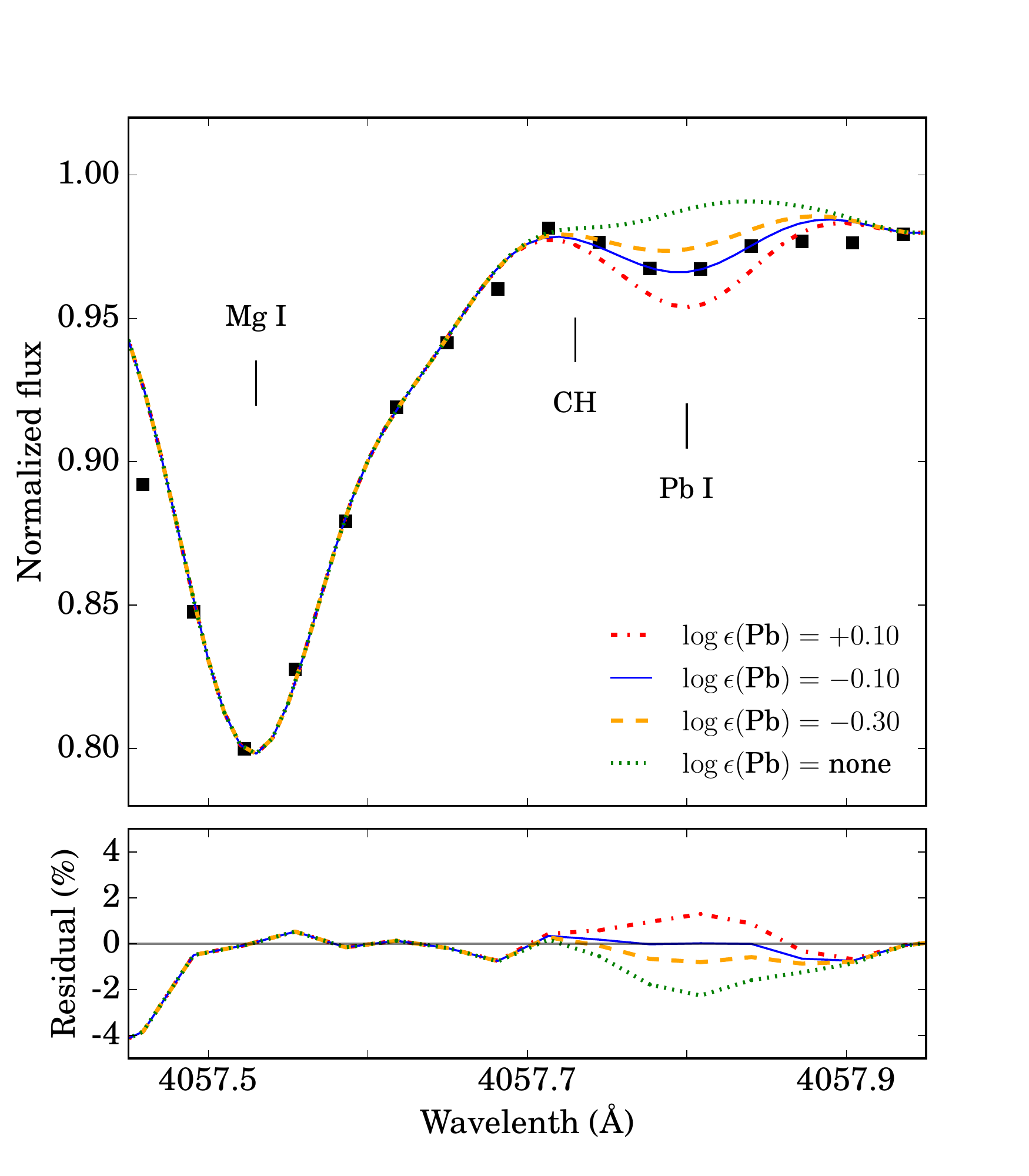}
\vfill
\end{minipage}
\caption{Observed spectra (squares) and synthesis (lines) for varying isotopic ratios
of \ion{Eu}{2} (left panel), and varying abundances of \ion{Eu}{2} (middle panel), 
and \ion{Pb}{1} (right panel).}
\label{fig:Eusyn_iso}
\end{figure*}

\paragraph{Strontium, Yttrium, Zirconium}
These three elements belong to the first \emph{r}-process peak, and are often
attributed to the weak \emph{r}-process \citep{wanajo2006}. Their abundances
are mostly determined from absorption lines in blue spectral regions, which can
be affected by the presence of carbon features in CEMP stars, which does not
apply to our current analysis. \citet{siqueira2014} have contrasted the
behavior of these elements in \emph{r}-I and \emph{r}-II stars, finding that
they are generally more enhanced in \emph{r}-I stars, and suggested the
possible existence of different nucleosynthesis pathways for these two
sub-classes of \emph{r}-process-enhanced stars.    


The Sr $\lambda$4077 is saturated, but $\lambda$4215 could be successfully
synthesized. The Sr line at $\lambda$4161 is a much weaker feature, and yields
a slightly higher abundance than the $\lambda$4215 line. The left panel of
Figure~\ref{fig:synfig} shows the spectral synthesis of the Sr $\lambda$4215
line for three different abundances, and also for the absence of Sr.

The Y $\lambda$4398, $\lambda$4682, $\lambda$4883, and
$\lambda$4900 lines are strong, well-isolated, and unsaturated. The
feature at $\lambda$4982 is weak, but its abundance agrees well with the
other four features. 

Although most features of Zr are weak in the spectrum of {\rii}, they are of
sufficient strength to extract an abundance estimate from spectral
synthesis; a total of six Zr features were used. 

The final adopted abundances for the first peak elements are
$\abund{Sr}{Fe}=+0.54$, $\abund{Y}{Fe}=+0.21$, and $\abund{Zr}{Fe}=+0.70$.

\paragraph{Barium, Lanthanum}
These elements constitute the second \emph{r}-process peak. The Ba lines on the
blue spectral range were mostly saturated, so only the three lines
at $\lambda$5853, $\lambda$6141, and $\lambda$6496 were
used to determine the overall Ba abundance. Hyperfine splitting was accounted
for in the spectral synthesis. The middle panel of Figure~\ref{fig:synfig}
shows the spectral synthesis of the Ba $\lambda$5853 line for three different
abundances, and also for the absence of Ba.  \emph{r}-process isotopic
fractions derived from solar abundances \citep{arlandini1999}. These
approximate the isotopic splitting of barium in {\rii} in order to produce a
more accurate synthetic fit. 

Six lines of \ion{La}{2} were identified, and their derived abundances 
agree within 0.3~dex.

Final abundances of the second-peak elements are $\abund{Ba}{Fe}=+0.83$ and 
$\abund{La}{Fe}=+1.05$.

\paragraph{Cerium, Praseodymium, Neodymium, Samarium}
A total of twelve \ion{Ce}{2} lines were used to determine the
abundance of cerium, more
features than for any other neutron-capture element in {\rii}. All Ce lines
agree with the adopted abundance of $\abund{Ce}{Fe}=+0.98$ within 0.2 dex.

Seven strong praseodymium lines were used to find the \ion{Pr}{2} abundance.
There are two \ion{Pr}{2} lines near the wings of the wide \ion{Ca}{2}~H feature
at $\lambda$3968, but abundances derived from these lines still agree with the
final abundance of $\abund{Pr}{Fe}=+1.30$.  Similarly, the feature at
$\lambda$4179 shows a blend with \ion{Nd}{2}, but still agrees well with the
adopted abundance.

There are many strong lines of neodymium, but several are blended with
absorption features of other elements. In total, eleven 
\ion{Nd}{2} features were used to determine the final abundance $\abund{Nd}{Fe}=+1.30$.

Five strong lines of samarium were used to determine the \ion{Sm}{2} abundance.
Although some lines showed a blend with other elements, all were taken into
account, and the derived abundances agree with the final abundance
$\abund{Sm}{Fe}=+1.42$ within 0.2 dex.

\paragraph{Europium} 
There are six europium lines with good agreement in their derived abundances,
yielding an average $\abund{Eu}{Fe}= +1.64$. The middle panel of
Figure~\ref{fig:Eusyn_iso} shows the spectral syntheses of the Eu $\lambda$4129
line for three different abundances, and also for the absence of Eu. Similarly to
Ba, the strong lines of Eu are sensitive to hyperfine splitting between the
isotopes $^{151}$Eu and $^{153}$Eu. With high-resolution spectroscopy, this
splitting has to be accounted for in spectral synthesis in order to fit the
\ion{Eu}{2} absorption features and measure the abundance. An isotopic ratio of
$^{151}\text{Eu}/^{153}\text{Eu}=0.902$ \citep[from solar \emph{r}-process
residuals;][]{arlandini1999} was used to approximate the effect of hyperfine
splitting on the \ion{Eu}{2} absorption features. 
%
%
For example, the left panel of
Figure~\ref{fig:Eusyn_iso} shows the effect of varying the isotopic fraction
$^{151}\text{Eu}/^{153}\text{Eu}$ at constant overall $\log\epsilon(\text{Eu})$
abundance. Clearly, only considering one isotope of Eu is not sufficient to
describe the line shape. However the specific isotopic fraction cannot be
measured this way. We only used it to calculate the \ion{Eu}{2} abundance.

\paragraph{Gadolinium, Terbium, Dysprosium, Holmium, Erbium} Gadolinium features
are often blended with neighboring lines or are located in the wings of
strong hydrogen features, making their abundance measurements difficult. Still,
all nine features agree within 0.2 dex of the adopted value of
$\abund{Gd}{Fe}=+1.47$.

Five lines of terbium were used to estimate the \ion{Tb}{2} abundance. The three
\ion{Tb}{2} lines at $\lambda$3702, $\lambda$3747, and $\lambda$3848 are blended
with other features. However, two clean features at $\lambda$3874 and
$\lambda$4002 yield similar abundances as those derived from the other three
lines. The final adopted abundance is $\abund{Tb}{Fe}=+1.38$.

The dysprosium abundances derived from eight \ion{Dy}{2} absorption features
show a spread of 0.24~dex, yielding abundances around either $\eps{Dy} = -0.42$
or $\eps{Dy} = -0.25$. The final abundance is taken to be $\abund{Dy}{Fe}=+1.48$
by averaging all eight lines.

Three features of holmium were used to estimate its abundance.  The
\ion{Ho}{2} lines at $\lambda$3810 and $\lambda$3890 in particular are
blended with Fe features. However, since the estimates for the three lines agree
within less than 0.1 dex, all were used for determining the final abundance,
$\abund{Ho}{Fe}=+1.40$.

The abundance of erbium was estimated from seven \ion{Er}{2} lines, all agreeing
to within 0.2 dex of the adopted value, $\abund{Er}{Fe}=+1.60$.

\paragraph{Thulium, Ytterbium, Lutetium, Hafnium}  %
Many \ion{Tm}{2} features are found in the blue ($\lambda<4000$ \AA) region of
the spectrum. Five lines were used, and their estimates are in good agreement,
yielding $\abund{Tm}{Fe}=+1.60$.

Only one strong \ion{Yb}{2} line can be measured in the spectrum of {\rii}.
It neighbors a blended Fe-Ni feature to the blue. Regardless, the line at
$\lambda$3694 was sufficiently strong to measure a ytterbium abundance with
confidence ($\abund{Yb}{Fe}=+1.56$). The right panel of Figure~\ref{fig:synfig}
shows the spectral synthesis of the Yb $\lambda$3694 line for three different
abundances, and also for the absence of Yb.
Since the \ion{Yb}{2} feature is well-described within a $\pm$0.2~dex
variation from the adopted values, an uncertainty of $\pm$0.2~dex is
assigned to the \ion{Yb}{2} abundance.

Similarly, only two \ion{Lu}{2} lines could be accurately fit with spectral
synthesis. Both lines appear far in the blue, around 3500\,{\AA}, near the edge
of the spectrum. The line at $\lambda$3472 appears blueward of a strong
\ion{Ni}{1} feature, so the line was fit by analyzing the asymmetry of the
blended feature and fitting the blue wing. The line at $\lambda$3507 is blended
with an iron feature, but its synthetic abundance agrees with that of the line
at $\lambda$3472\,\AA, yielding a lutetium abundance of $\abund{Lu}{Fe}=+1.39$.

Four hafnium lines were used to etimate the final abundance of \ion{Hf}{2}. The
cleanest feature at $\lambda$3719 yielded the highest abundance,
$\eps{Hf}=-0.5$. The two features at $\lambda$3793 and 3918 are uncertain, but
agree with each other. The feature at $\lambda$4093 has the lowest derived
abundance, $\eps{Hf} = -0.79$. The adopted hafnium abundance, from an average of
all four lines, is $\abund{Hf}{Fe} = +1.40$.

\paragraph{Osmium and Iridium} 
Osmium and iridium represent the third \emph{r}-process peak. 

Three lines of \ion{Os}{1} were used. The individual abundances calculated from
these lines disagree (range of 0.45 dex), but it is still clear that the
features are present. We obtain a final estimate of $\abund{Os}{Fe} = +1.60$ 
by averaging the three individual measurements.

Only the line at $\lambda$3800 could be used to estimate the iridium
abundance. This feature appears in a crowded part of the spectrum, but the line
is unblended, and could be well-fit by spectral synthesis. The adopted abundance
of \ion{Ir}{1} from this line is $\abund{Ir}{Fe}=+1.51$.

\paragraph{Lead}  
This third-peak element is typically largely produced by the \emph{s}-process
\citep{travaglio2001} in metal-poor stars. However, since {\rii} exhibits no
\emph{s}-process enhancement, and its neutron-capture elements likely result
from only \emph{r}-process events, the presence of Pb in {\rii} is not
indicative of \emph{s}-process enrichment. Rather, \citet{wanajo2002} call
attention to the importance of the third-peak element Pb for understanding the
nature of the \emph{r}-process, since it is mainly synthesized by the same
$\alpha$-decay chains as Th and U.

Only the line at $\lambda$4057 was used to obtain the \ion{Pb}{1}
abundance. The right panel of Figure~\ref{fig:Eusyn_iso} shows the spectral
synthesis of this line for three different abundances, and also for the absence
of Pb. The final abundance is $\abund{Pb}{Fe}=+1.06$.

\begin{figure*}[!ht]
\centering
\begin{minipage}{0.32\textwidth}
\centering
\includegraphics[width=\textwidth]{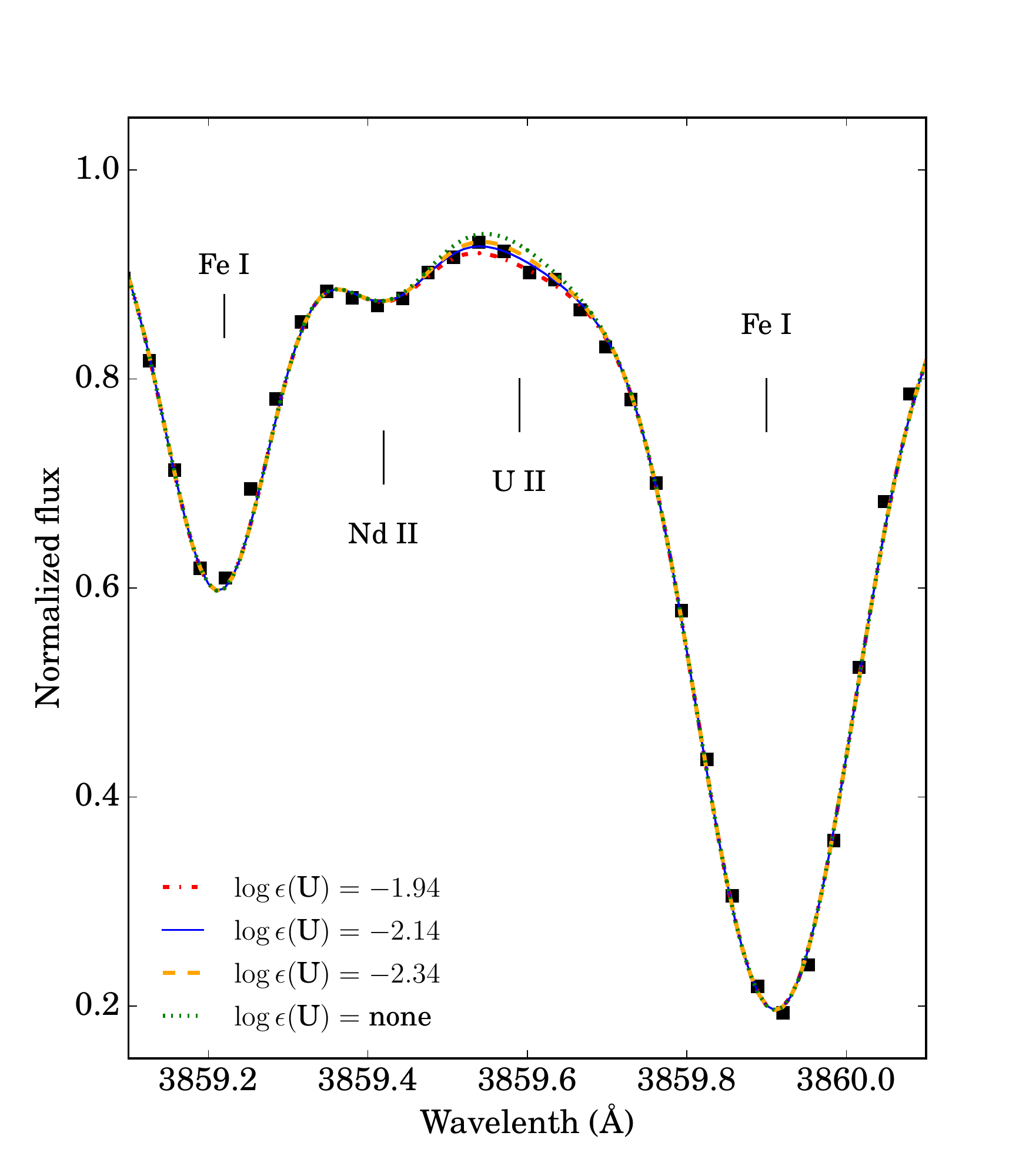}
\end{minipage}%
\begin{minipage}{0.32\textwidth}
\includegraphics[width=\textwidth]{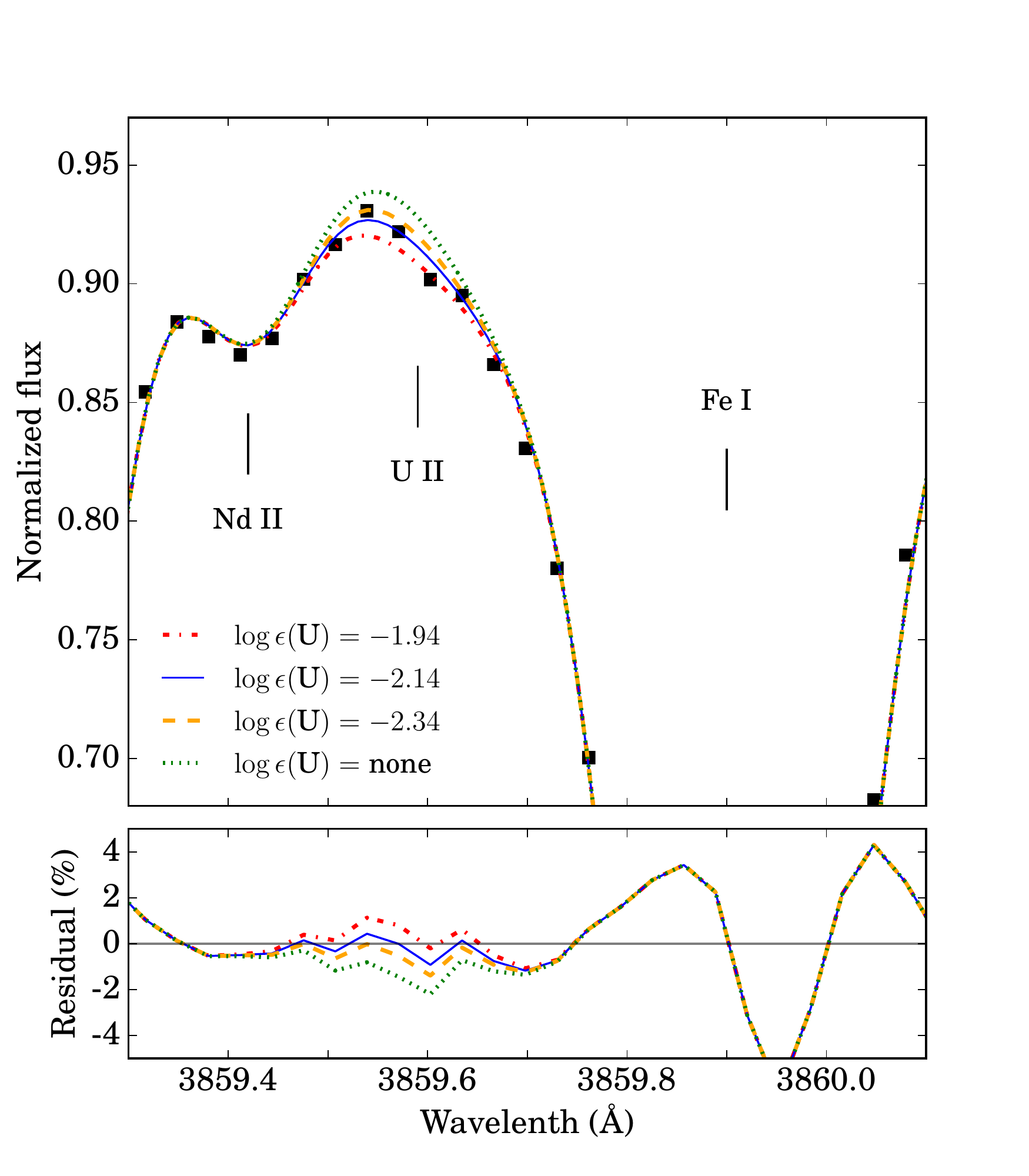}
\vfill
\end{minipage}
\begin{minipage}{0.32\textwidth}
\includegraphics[width=\textwidth]{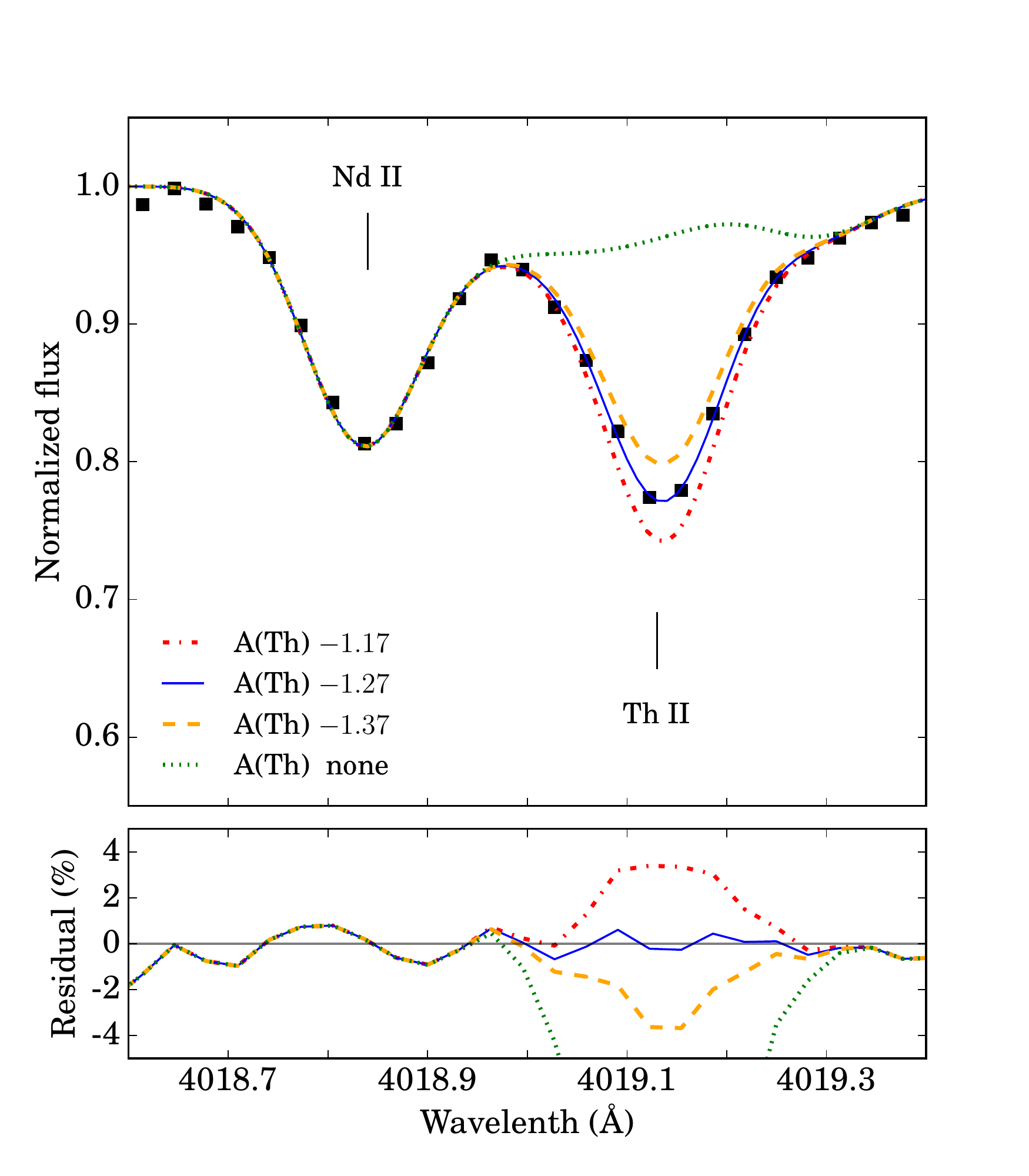}
\vfill
\end{minipage}
\caption{Observed spectra (squares) and synthesis (lines) of varying abundances
of \ion{U}{2} $\lambda$3859 (left and middle panels) and \ion{Th}{2}
$\lambda$4019 (right panel).}
\label{fig:uranium}
\end{figure*}

\paragraph{Thorium and Uranium}
Thorium and uranium are radioactive actinides, and the heaviest observable
elements in a stellar spectrum. These can only be synthesized in an
\emph{r}-process event. Furthermore, their presence allow the calculation of
stellar ages (see Section \ref{secage}).

Three lines of \ion{Th}{2} were used to determine the final abundance. The
feature at $\lambda$4019 (right panel of Figure~\ref{fig:uranium}) is the
strongest, and blended with CH, Ni, and Pr features, which were accounted for in
the synthesis.  The other two Th features at $\lambda$4086 and $\lambda$4094
agree well with the adopted abundance, $\abund{Th}{Fe} = +1.65$.

Only one \ion{U}{2} feature could be measured with accuracy. The
$\lambda$3859 line appears at the far edge of a strong iron line, between
a \ion{Nd}{2} and a CN feature. After the abundances of neodymium and carbon are
well-determined, the uranium feature was fit, and an abundance of
$\abund{U}{Fe}=+1.31$ fits the data most accurately (see left and middle panels
of Figure~\ref{fig:uranium}). From inspection of this figure, it is clear that a
higher S/N spectrum would be useful to better constrain this
determination. 

\section{Discussion}
 \label{seccomp}

 \subsection{Heavy-Element Pattern for \protect\rii}

\begin{figure*}[!ht]
\epsscale{1.05}
\plotone{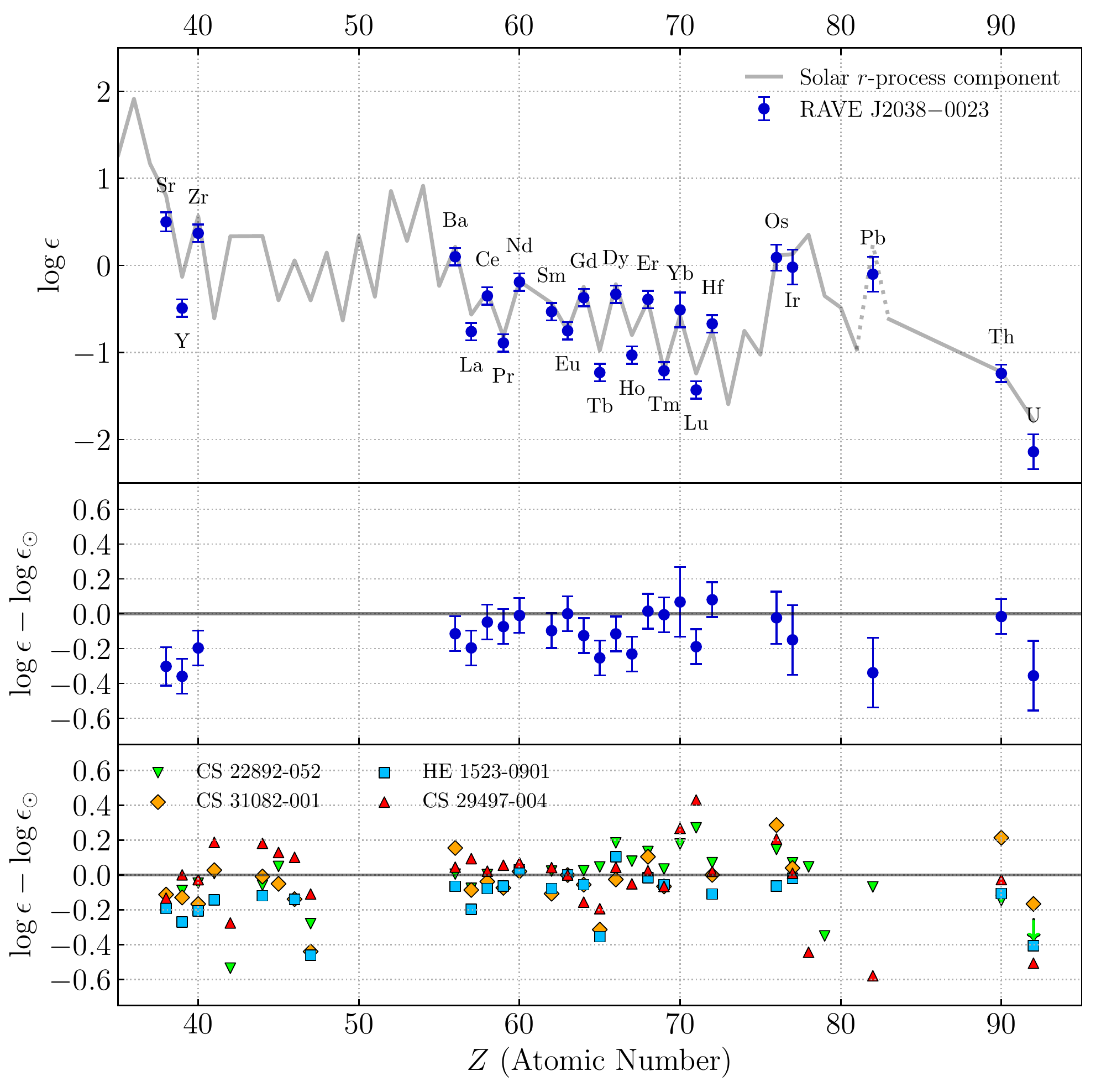}
\caption{Top: Full neutron-capture elemental-abundance pattern for {\rii}
compared with the Solar System \emph{r}-process component
\citep{arlandini1999}, normalized to Eu. Middle: Difference between {\rii}
abundances and the Solar System \emph{r}-process component, normalized to Eu. 
Bottom: Same as middle panel, for the stars: 
CS~22892$-$052 \citep{sneden2008},
CS~29497$-$004 \citep{hill2016}, 
CS~31082$-$001 \citep{hill2002}, and
HE~1523$-$0901 \citep{frebel2007}.}
\label{fig:pattern}
\end{figure*}

The top panel of Figure \ref{fig:pattern} compares the measured neutron-capture
elemental-abundance pattern with the scaled Solar System \emph{r}-process
pattern, normalized to the Eu abundance. The abundances of {\rii} agree well,
and deviations from the Solar System \emph{r}-process pattern (shown in the
middle panel of Figure~\ref{fig:pattern}, normalized to Eu) indicate a
suppression in the first \emph{r}-process peak (elements Sr, Y, and Zr). 

The first-peak elements are of particular
interest, as they have been suggested by \citet{siqueira2014} and others to be
associated with production by the weak, rather than the main, \emph{r}-process
\citep[perhaps by neutrino-driven winds in core-collapse SNe -][]{arcones2007,
wanajo2007,arcones2013}. The slight under-abundance of the first-peak elements
in {\rii} may support the argument for multiple \emph{r}-process sites in which
a weak \emph{r}-process efficiently produces first-peak neutron-capture
elements, but cannot robustly synthesize elements beyond the second peak.
Furthermore, \cite{siqueira2014} have also presented evidence that the
first-peak elements for moderately \emph{r}-process-enhanced (\emph{r}-I) stars
generally exceed the levels found for \emph{r}-II stars, and that this may
indicate the operation of different nucleosynthesis pathways for these classes
of stars.

\begin{figure*}[!ht]
\epsscale{1.05}
\plotone{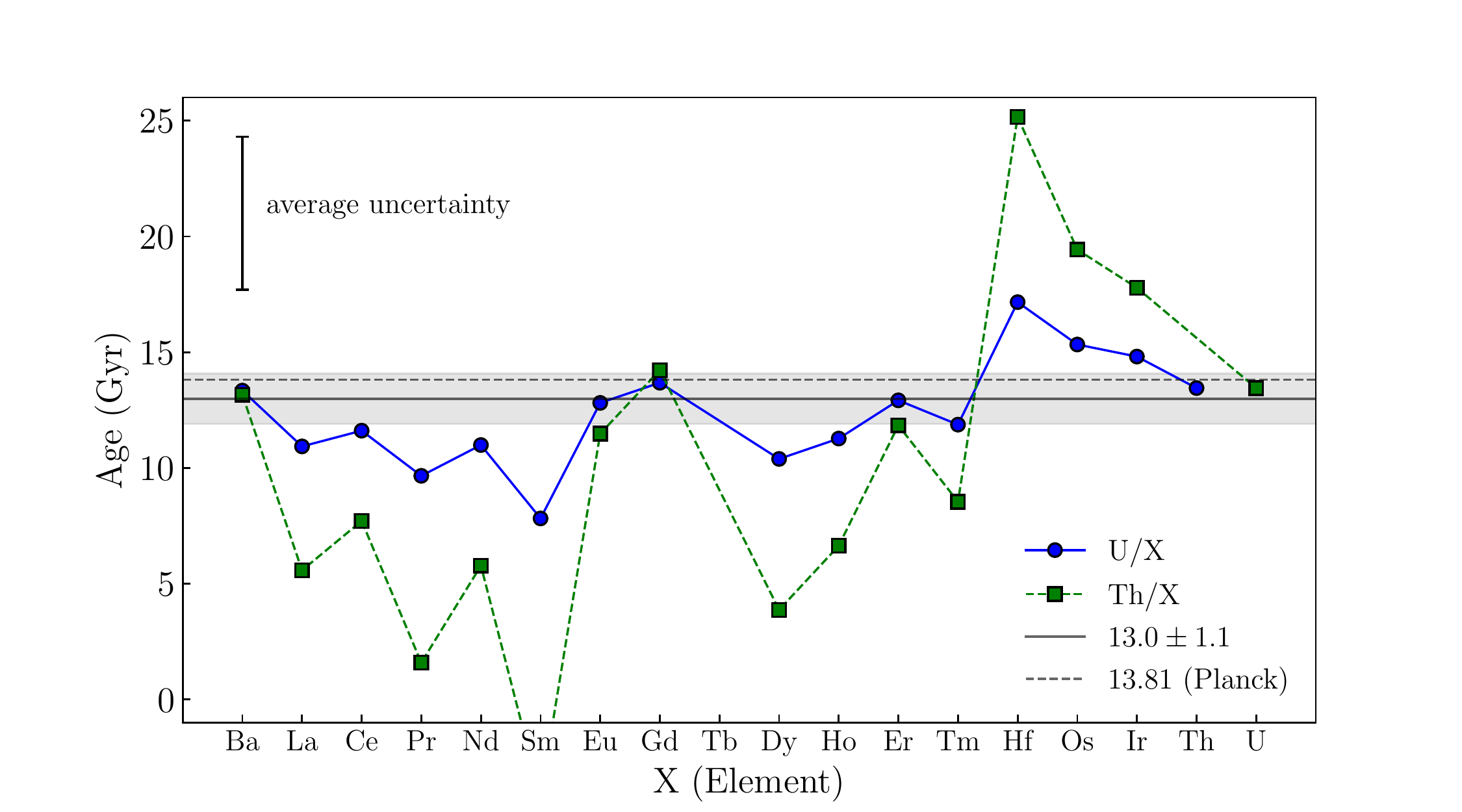}
\caption{Radioactive-decay ages using PRs from \cite{hill2016} for U/X (blue,
solid, circles) and Th/X (green, dashed, squares). 
The solid horizontal line marks the mean age for \protect\rii\
(see text for details), the shaded area represents its uncertainty, and 
the dashed horizontal line is the age of the Universe 
\citep{planck2016xiii}. Also shown on the top left is a typical error bar for
the individual age determinations. The production ratios and derived ages for individual
chronometer pairs are listed Table \ref{tab:ages}.
\explain{small edits made to the Figure, to reflect changes in the text.}}
\label{fig:ages} 
\end{figure*}

Among the heaviest stable neutron-capture elements, it is worth noting that Pb
is most discrepant from the scaled Solar System \emph{r}-process pattern, for both
\rii\ and CS~29407$−$004. This is due to the fact the the solar
\emph{r}-process pattern is derived from subtracting the \emph{s}-process component of the
total solar abundance pattern. The \emph{s}-process at higher metallicity produces
less Pb compared to other neutron-capture elements. This leads to an
overestimated \emph{r}-process Pb contribution when comparing a scaled solar r-process
pattern and a low-metallicity star, as seen in Figure~\ref{fig:pattern}. For a more accurate
comparison, a ``low-metallicity" solar-process pattern would have to be derived.
\rii\ has a \abund{Pb}{Eu}$ = -0.68$, which is consistent with the production by
the \emph{r}-process at low metallicities (see \citealt{roederer2010a} and
Figure~15 of \citealt{placco2013}).

The bottom panel of Figure~\ref{fig:pattern} compares the residual abundances of 
four other \emph{r}-II stars with reported measurements (and upper limits) of U.
Although there appears to be some
small differences in the derived abundances between the U stars, their patterns
largely agree within the uncertainties. Even among the \emph{r}-II stars, there
appears to be some real scatter in the first-peak elements Sr, Y, and Zr, with
{\rii} being generally lower than the other U stars. For Th and U, {\rii}
appears commensurate with CS~29407$-$004 and HE~1523$-$0901, which are not
actinide-boost\footnote{stars with enhancements in Th and U abundance ratios
relative to the rare earth elements \citep{schatz2002,roederer2009}.}
stars, and all three are lower than the one U star known to exhibit an actinide
boost, CS~31082$-$001. The actinide boost has also been recognized in
HE~1219$-$0312, another \emph{r}-II star with measurable Th, but lacking a
detectable uranium feature \citep{barklem2005}, given its faint magnitude. The
over-abundance of the actinides in some \emph{r}-II stars might suggest
different \emph{r}-process formation scenarios involving one or multiple
\emph{r}-process from sources, such as a high-entropy wind from SNII and the
ejecta from neutron star mergers \citep[see][for further
details]{mashonkina2010}.

\subsection{Age Determinations} \label{secage}

The presence of radioactive elements Th and U in {\rii} allows for the
determination of the star's age (or more correctly, the time that has passed
since the production of these elements) through radioactive decay dating.
Radioactive decay ages are estimated by measuring the relative abundances of
long-lived radioactive elements (i.e., $^{232}$Th: $t_{1/2}= 14.0$ Gyr, and
$^{238}$U: $t_{1/2}=4.5$ Gyr) to stable elements (i.e., the ratios Th/X and
U/X, where X is a stable element) or the ratio between the radioactive elements
themselves (Th/U).

To make use of radioactive decay dating, a set of initial production ratios
(PRs: $\text{Th/X}_0$, $\text{U/X}_0$, and $\text{Th/U}_0$) must be estimated.
For the present work, we use PRs from (i) the \emph{r}-process waiting-point
calculations of \cite{schatz2002} and from (ii) \cite{hill2016} based on the
high-entropy wind model of \cite{farouqi2010}. With PRs in hand, the ages $t$
are calculated as follows: 
\begin{eqnarray}
	t =& 46.67\mbox{ Gyr } \left[\log\epsilon\left(\text{Th/X}\right)_0 - \log\epsilon\left(\text{Th/X}\right)_{\text{obs}}\right],\label{eqn:thage}\\
	t =& 14.84\mbox{ Gyr } \left[\log\epsilon\left(\text{U/X}\right)_0 - \log\epsilon\left(\text{U/X}\right)_{\text{obs}}\right],\label{eqn:uage}
\end{eqnarray}
and
\begin{eqnarray}
	t =& 21.80\mbox{ Gyr } \left[\log\epsilon\left(\text{U/Th}\right)_0 - \log\epsilon\left(\text{U/Th}\right)_{\text{obs}}\right],\label{eqn:uthage}
\end{eqnarray}

\noindent where $\log\epsilon\left(\text{Th/X}\right)_0$ is the initial PR corresponding
to element formation at $t=0$, and
$\log\epsilon\left(\text{Th/X}\right)_{\text{obs}}$ is the observed ratio after
the radioactive elements Th and U have decayed for a time $t$. The half-lives
of Th and U are contained in the constant. Table \ref{tab:ages} lists the PRs
used in the above equations and the ages derived from these abundance ratios.
The calculated U/X and Th/X ages using the PRs from \cite{hill2016} are shown
in Figure \ref{fig:ages}. 
The solid horizontal line marks the mean age for \rii\ (see details
below), and the dashed horizontal line is the age of the Universe, determined by
the Planck mission \citep{planck2016xiii}.

Arithmetic means were taken for all four sets of ages: U/X and Th/X, each using
the two different sets of PRs described above. All U/X and Th/X ages agree
within $2\sigma$ of their arithmetic means except for ages calculated from Hf
ratios in the \cite{schatz2002} model (see Table \ref{tab:ages}). 
The small uncertainty on the Hf abundance suggests that the discrepancy between
X/Hf and other ages is driven by the predicted production ratio for this
chronometer pair by both \emph{r}-process models considered above. The same
inconsistency of X/Hf ages was also noticed by \cite{hill2016} in their analysis
of the U star CS~29497$-$004.

The uncertainties in Table \ref{tab:ages} reflect only the propagated error
from the abundance measurement uncertainty; systematic errors from the model
atmosphere parameters as well as uncertainties from the \emph{r}-process models
considered here were not included. Since only one \ion{U}{2} feature could be
measured, the uncertainty associated with the \ion{U}{2} abundance is set to
0.2~dex. The syntheses in Figure \ref{fig:uranium} show the best-fit abundance
as well as $\pm 0.2$ dex from the best fit. Since the feature is well-described
within these limits, an uncertainty of 0.2 dex is suitable for the uranium
abundance. 

Although the uncertainty on the uranium abundance is larger than that of
thorium (0.2 and 0.04 dex, respectively), the individual U/X ages vary much
less than the Th/X ages. This apparent contradiction results primarily from the
longer half-life of $^{232}$Th (and therefore the larger constant in Equation
\ref{eqn:thage}), causing Th/X ages to be much more sensitive to variations on
the measured Th/X abundance ratios. On the other hand, U/X ages---albeit
carrying large uncertainties---agree with the expected ages of VMP/EMP stars.

From Figure~\ref{fig:ages}, it is interesting to note that the ages
calculated for Ba, Eu, Gd, and Er, using both Th/X and U/X ratios, agree with
each other and with the Th/U age of $13.4$ Gyr, using the PRs of
\citet{hill2016}.  The mean age for Th/X, using Ba, Eu, Gd, and Er, is $12.7\pm
1.6$ Gyr, while the mean age for U/X with the same four elements is $13.2\pm
1.5$ Gyr. The fact that these elements present a much smaller scatter between
the Th/X and U/X ratios suggests that these are  more likely to represent a 
realistic age for \rii. Then, by averaging Th/X and U/X ratios for Ba, Eu, Gd,
and Er, and Th/U \citep[with PRs from][]{hill2016}, we adopt an age of $13.0\pm
1.1$ Gyr for \rii.

\section{Conclusions}
\label{final}

We have presented results for the first \emph{r}-II star discovered from the
RAVE survey, {\rii}, and only the fourth \emph{r}-II star with measured Th and
U. This star was first identified as a metal-poor candidate from the RAVE DR4, 
and then followed-up with medium- and high-resolution spectroscopy with the Mayall
and Magellan telescopes, respectively.

A detailed high-resolution abundance analysis reveals that the chemical
abundance pattern of {\rii} nearly duplicates the scaled Solar System
\emph{r}-process pattern, similar to the other three known U stars and other
\emph{r}-II stars.  With measured abundances for the actinides thorium and
uranium, we were able to determine radioactive-decay ages for {\rii} from Th/X
and U/X abundance ratios, using initial production ratios from an
\emph{r}-process high-entropy wind model. The estimated  age for \rii\
(13.0$\pm$1.1~Gyr) is consistent with expectations for the epoch in which VMP/EMP
stars formed.  We note that the yields of a neutron star merger r-process may
differ from the standard scenarios considered here. We plan to update our age
estimates as realistic yields from these events become available.

We are presently extending our effort to identify large numbers of \emph{r}-II
(and \emph{r}-I) stars, based on medium-resolution spectroscopy of a large
number ($\sim 2000-2500$) of bright targets with [Fe/H] $< -2.0$ from a number
of sources, in addition to RAVE.  High-resolution spectroscopic follow-up of
these targets, already underway, should identify on the order of 75 new
\emph{r}-II (and 350 new \emph{r}-I) stars. This would provide a sufficiently
large sample to carry out detailed tests of the likely astrophysical site(s) of
the production of the \emph{r}-process elements in the early Galaxy, and tests
of the association of \emph{r}-II (and \emph{r}-I) stars in the field with
particular environments, such as the ultra-faint and canonical dwarf galaxies
in which similar stars have been previously identified \citep[see][and
references therein]{hansen2017}.

\acknowledgments

We thank Ian Roederer for providing the high-resolution spectra of
CS~22892$-$052 used in Figure~\ref{highres}. The authors acknowledge partial
support for this work from grant PHY 14-30152; Physics Frontier Center/JINA
Center for the Evolution of the Elements (JINA-CEE), awarded by the US National
Science Foundation.  A.F. acknowledges support from NSF CAREER grant
AST-1255160. 

\software{gnuplot \citep{gnuplot},
IRAF \citep{tody1986,tody1993},
Matplotlib \citep{matplotlib},
MIKE data reduction pipeline \citep{kelson2003},
MOOG \citep[July 2014 version;][]{sneden1973},
n-SSPP \citep{beers2014},
SMH \citep{casey2014}.}

\bibliographystyle{apj}

\startlongtable
\begin{deluxetable}{@{}l@{}r@{}r@{}r@{}r@{}r@{}}
\tabletypesize{\scriptsize}
\tablewidth{0pc}
\tablecaption{\label{eqw} Equivalent-Width Measurements for \protect\rii}
\tablehead{
\colhead{Ion}&
\colhead{$\lambda$}&
\colhead{$\chi$} &
\colhead{$\log\,gf$}&
\colhead{$W$}&
\colhead{$\log\epsilon$\,(X)}\\
\colhead{}&
\colhead{({\AA})}&
\colhead{(eV)} &
\colhead{}&
\colhead{(m{\AA})}&
\colhead{}}
\startdata
      Na I   &  5889.950  &  0.00  &   0.108  &  171.69  &   3.72 \\ 
      Mg I   &  3829.355  &  2.71  &  $-$0.208  &  153.47  &   5.08 \\ 
      Mg I   &  3832.304  &  2.71  &   0.270  &  192.31  &   4.97 \\ 
      Mg I   &  3986.753  &  4.35  &  $-$1.030  &   21.60  &   5.10 \\ 
      Mg I   &  4057.505  &  4.35  &  $-$0.890  &   23.68  &   5.00 \\ 
      Mg I   &  4167.271  &  4.35  &  $-$0.710  &   35.18  &   5.05 \\ 
      Mg I   &  4571.096  &  0.00  &  $-$5.688  &   62.43  &   5.13 \\ 
      Mg I   &  4702.990  &  4.33  &  $-$0.380  &   60.66  &   5.02 \\ 
      Mg I   &  5172.684  &  2.71  &  $-$0.450  &  182.51  &   5.04 \\ 
      Mg I   &  5183.604  &  2.72  &  $-$0.239  &  208.96  &   5.08 \\ 
      Mg I   &  5528.405  &  4.34  &  $-$0.498  &   60.46  &   5.06 \\ 
      Al I   &  3961.520  &  0.01  &  $-$0.340  &  122.10  &   3.09 \\ 
      Si I   &  3905.523  &  1.91  &  $-$1.092  &  192.86  &   5.17 \\ 
      Si I   &  4102.936  &  1.91  &  $-$3.140  &   74.50  &   5.21 \\ 
      Ca I   &  4226.730  &  0.00  &   0.244  &  187.14  &   3.47 \\ 
      Ca I   &  4283.010  &  1.89  &  $-$0.224  &   36.01  &   3.51 \\ 
      Ca I   &  4318.650  &  1.89  &  $-$0.210  &   39.49  &   3.56 \\ 
      Ca I   &  4425.440  &  1.88  &  $-$0.358  &   30.44  &   3.50 \\ 
      Ca I   &  4455.890  &  1.90  &  $-$0.530  &   24.40  &   3.55 \\ 
      Ca I   &  5588.760  &  2.52  &   0.210  &   34.30  &   3.64 \\ 
      Ca I   &  5594.468  &  2.52  &   0.097  &   31.59  &   3.69 \\ 
      Ca I   &  5598.487  &  2.52  &  $-$0.087  &   20.71  &   3.63 \\ 
      Ca I   &  5857.450  &  2.93  &   0.230  &   15.88  &   3.65 \\ 
      Ca I   &  6102.720  &  1.88  &  $-$0.790  &   23.66  &   3.61 \\ 
      Ca I   &  6122.220  &  1.89  &  $-$0.315  &   52.46  &   3.67 \\ 
      Ca I   &  6162.170  &  1.90  &  $-$0.089  &   63.69  &   3.63 \\ 
      Ca I   &  6439.070  &  2.52  &   0.470  &   48.99  &   3.58 \\ 
     Sc II   &  4314.083  &  0.62  &  $-$0.100  &   87.86  &   0.27 \\ 
     Sc II   &  4324.998  &  0.59  &  $-$0.440  &   70.00  &   0.18 \\ 
     Sc II   &  4400.389  &  0.61  &  $-$0.540  &   63.60  &   0.17 \\ 
     Sc II   &  4415.544  &  0.59  &  $-$0.670  &   59.26  &   0.19 \\ 
     Sc II   &  5031.010  &  1.36  &  $-$0.400  &   28.79  &   0.23 \\ 
     Sc II   &  5526.785  &  1.77  &   0.020  &   24.65  &   0.17 \\ 
     Sc II   &  5657.907  &  1.51  &  $-$0.600  &   16.86  &   0.26 \\ 
     Sc II   &  6604.578  &  1.36  &  $-$1.310  &    5.94  &   0.21 \\ 
      Ti I   &  3989.760  &  0.02  &  $-$0.062  &   60.52  &   2.17 \\ 
      Ti I   &  3998.640  &  0.05  &   0.010  &   59.89  &   2.12 \\ 
      Ti I   &  4533.249  &  0.85  &   0.532  &   44.80  &   2.12 \\ 
      Ti I   &  4534.780  &  0.84  &   0.336  &   32.51  &   2.07 \\ 
      Ti I   &  4535.567  &  0.83  &   0.120  &   24.87  &   2.12 \\ 
      Ti I   &  4656.470  &  0.00  &  $-$1.289  &   12.26  &   2.12 \\ 
      Ti I   &  4681.910  &  0.05  &  $-$1.015  &   18.01  &   2.09 \\ 
      Ti I   &  4981.730  &  0.84  &   0.560  &   51.75  &   2.12 \\ 
      Ti I   &  4991.070  &  0.84  &   0.436  &   43.39  &   2.10 \\ 
      Ti I   &  4999.500  &  0.83  &   0.306  &   40.33  &   2.16 \\ 
      Ti I   &  5039.960  &  0.02  &  $-$1.130  &   17.85  &   2.12 \\ 
      Ti I   &  5210.390  &  0.05  &  $-$0.828  &   29.14  &   2.11 \\ 
     Ti II   &  3813.394  &  0.61  &  $-$2.020  &   59.03  &   2.19 \\ 
     Ti II   &  4025.120  &  0.61  &  $-$1.980  &   65.52  &   2.18 \\ 
     Ti II   &  4053.829  &  1.89  &  $-$1.210  &   26.32  &   2.19 \\ 
     Ti II   &  4161.527  &  1.08  &  $-$2.160  &   28.86  &   2.18 \\ 
     Ti II   &  4290.219  &  1.16  &  $-$0.930  &   84.74  &   2.11 \\ 
     Ti II   &  4300.049  &  1.18  &  $-$0.490  &  102.20  &   2.12 \\ 
     Ti II   &  4330.723  &  1.18  &  $-$2.060  &   30.50  &   2.20 \\ 
     Ti II   &  4337.914  &  1.08  &  $-$0.960  &   89.61  &   2.14 \\ 
     Ti II   &  4394.059  &  1.22  &  $-$1.780  &   44.15  &   2.21 \\ 
     Ti II   &  4395.031  &  1.08  &  $-$0.540  &  109.36  &   2.17 \\ 
     Ti II   &  4395.839  &  1.24  &  $-$1.930  &   31.59  &   2.15 \\ 
     Ti II   &  4399.765  &  1.24  &  $-$1.190  &   70.75  &   2.13 \\ 
     Ti II   &  4417.714  &  1.17  &  $-$1.190  &   74.65  &   2.12 \\ 
     Ti II   &  4418.331  &  1.24  &  $-$1.970  &   29.85  &   2.15 \\ 
     Ti II   &  4443.801  &  1.08  &  $-$0.720  &  102.84  &   2.15 \\ 
     Ti II   &  4444.554  &  1.12  &  $-$2.240  &   25.48  &   2.18 \\ 
     Ti II   &  4450.482  &  1.08  &  $-$1.520  &   62.62  &   2.09 \\ 
     Ti II   &  4464.448  &  1.16  &  $-$1.810  &   45.36  &   2.17 \\ 
     Ti II   &  4468.517  &  1.13  &  $-$0.600  &  107.79  &   2.20 \\ 
     Ti II   &  4470.853  &  1.17  &  $-$2.020  &   35.00  &   2.21 \\ 
     Ti II   &  4501.270  &  1.12  &  $-$0.770  &   98.75  &   2.13 \\ 
     Ti II   &  4533.960  &  1.24  &  $-$0.530  &  101.99  &   2.10 \\ 
     Ti II   &  4563.770  &  1.22  &  $-$0.960  &   89.70  &   2.21 \\ 
     Ti II   &  4571.971  &  1.57  &  $-$0.320  &   96.06  &   2.15 \\ 
     Ti II   &  4589.915  &  1.24  &  $-$1.790  &   43.05  &   2.19 \\ 
     Ti II   &  4657.200  &  1.24  &  $-$2.240  &   19.14  &   2.13 \\ 
     Ti II   &  4708.662  &  1.24  &  $-$2.340  &   18.37  &   2.20 \\ 
     Ti II   &  4779.979  &  2.05  &  $-$1.370  &   17.26  &   2.18 \\ 
     Ti II   &  4805.089  &  2.06  &  $-$1.100  &   27.86  &   2.18 \\ 
     Ti II   &  5129.156  &  1.89  &  $-$1.240  &   31.65  &   2.15 \\ 
     Ti II   &  5185.902  &  1.89  &  $-$1.490  &   22.26  &   2.19 \\ 
     Ti II   &  5336.786  &  1.58  &  $-$1.590  &   36.34  &   2.18 \\ 
     Ti II   &  5381.021  &  1.57  &  $-$1.920  &   21.86  &   2.19 \\ 
     Ti II   &  5418.768  &  1.58  &  $-$2.000  &   17.70  &   2.17 \\ 
      V II   &  3951.960  &  1.48  &  $-$0.784  &   30.75  &   1.28 \\ 
      V II   &  4035.622  &  1.79  &  $-$0.767  &   18.53  &   1.33 \\ 
      Cr I   &  3578.680  &  0.00  &   0.420  &   97.67  &   2.38 \\ 
      Cr I   &  4254.332  &  0.00  &  $-$0.114  &  108.51  &   2.45 \\ 
      Cr I   &  4274.800  &  0.00  &  $-$0.220  &  105.31  &   2.46 \\ 
      Cr I   &  4545.950  &  0.94  &  $-$1.370  &    8.53  &   2.38 \\ 
      Cr I   &  4600.752  &  1.00  &  $-$1.260  &   11.89  &   2.50 \\ 
      Cr I   &  4616.137  &  0.98  &  $-$1.190  &   13.15  &   2.45 \\ 
      Cr I   &  4626.188  &  0.97  &  $-$1.320  &   10.36  &   2.45 \\ 
      Cr I   &  4646.150  &  1.03  &  $-$0.740  &   27.92  &   2.46 \\ 
      Cr I   &  4651.280  &  0.98  &  $-$1.460  &    7.80  &   2.46 \\ 
      Cr I   &  4652.158  &  1.00  &  $-$1.030  &   15.28  &   2.39 \\ 
      Cr I   &  5206.040  &  0.94  &   0.020  &   82.59  &   2.49 \\ 
      Cr I   &  5296.690  &  0.98  &  $-$1.360  &   10.83  &   2.44 \\ 
      Cr I   &  5298.280  &  0.98  &  $-$1.140  &   17.81  &   2.47 \\ 
      Cr I   &  5345.800  &  1.00  &  $-$0.950  &   24.19  &   2.46 \\ 
      Cr I   &  5348.310  &  1.00  &  $-$1.210  &   13.40  &   2.42 \\ 
      Cr I   &  5409.770  &  1.03  &  $-$0.670  &   38.95  &   2.50 \\ 
      Mn I   &  4030.753  &  0.00  &  $-$0.480  &   98.98  &   1.98 \\ 
      Mn I   &  4033.062  &  0.00  &  $-$0.618  &   95.43  &   2.01 \\ 
      Mn I   &  4034.483  &  0.00  &  $-$0.811  &   89.39  &   2.03 \\ 
      Mn I   &  4041.357  &  2.11  &   0.285  &   23.28  &   2.02 \\ 
      Mn I   &  4754.048  &  2.28  &  $-$0.086  &    9.47  &   2.00 \\ 
      Mn I   &  4783.432  &  2.30  &   0.042  &   11.95  &   2.00 \\ 
      Mn I   &  4823.528  &  2.32  &   0.144  &   14.06  &   2.00 \\ 
      Fe I   &  3608.859  &  1.01  &  $-$0.090  &  149.26  &   4.48 \\ 
      Fe I   &  3767.192  &  1.01  &  $-$0.390  &  145.10  &   4.48 \\ 
      Fe I   &  3787.880  &  1.01  &  $-$0.838  &  125.95  &   4.60 \\ 
      Fe I   &  3815.840  &  1.48  &   0.237  &  155.99  &   4.49 \\ 
      Fe I   &  3840.438  &  0.99  &  $-$0.497  &  157.67  &   4.61 \\ 
      Fe I   &  3841.048  &  1.61  &  $-$0.044  &  132.50  &   4.58 \\ 
      Fe I   &  3845.169  &  2.42  &  $-$1.390  &   27.87  &   4.41 \\ 
      Fe I   &  3846.800  &  3.25  &  $-$0.020  &   52.97  &   4.60 \\ 
      Fe I   &  3849.967  &  1.01  &  $-$0.863  &  118.69  &   4.38 \\ 
      Fe I   &  3863.741  &  2.69  &  $-$1.430  &   18.18  &   4.51 \\ 
      Fe I   &  3865.523  &  1.01  &  $-$0.950  &  119.79  &   4.48 \\ 
      Fe I   &  3867.216  &  3.02  &  $-$0.450  &   42.76  &   4.51 \\ 
      Fe I   &  3885.510  &  2.42  &  $-$1.090  &   34.42  &   4.24 \\ 
      Fe I   &  3886.282  &  0.05  &  $-$1.080  &  179.35  &   4.55 \\ 
      Fe I   &  3895.656  &  0.11  &  $-$1.668  &  136.17  &   4.56 \\ 
      Fe I   &  3902.946  &  1.56  &  $-$0.442  &  119.56  &   4.62 \\ 
      Fe I   &  3920.258  &  0.12  &  $-$1.734  &  135.98  &   4.60 \\ 
      Fe I   &  3977.741  &  2.20  &  $-$1.120  &   66.77  &   4.68 \\ 
      Fe I   &  4001.661  &  2.18  &  $-$1.900  &   20.70  &   4.40 \\ 
      Fe I   &  4005.242  &  1.56  &  $-$0.583  &  114.09  &   4.56 \\ 
      Fe I   &  4007.272  &  2.76  &  $-$1.280  &   22.20  &   4.52 \\ 
      Fe I   &  4032.627  &  1.49  &  $-$2.380  &   36.85  &   4.40 \\ 
      Fe I   &  4044.609  &  2.83  &  $-$1.220  &   21.78  &   4.52 \\ 
      Fe I   &  4045.812  &  1.49  &   0.284  &  176.21  &   4.56 \\ 
      Fe I   &  4058.217  &  3.21  &  $-$1.110  &   12.93  &   4.58 \\ 
      Fe I   &  4062.441  &  2.85  &  $-$0.860  &   37.19  &   4.53 \\ 
      Fe I   &  4063.594  &  1.56  &   0.062  &  148.94  &   4.53 \\ 
      Fe I   &  4067.978  &  3.21  &  $-$0.470  &   31.54  &   4.45 \\ 
      Fe I   &  4070.769  &  3.24  &  $-$0.790  &   18.86  &   4.50 \\ 
      Fe I   &  4071.738  &  1.61  &  $-$0.008  &  146.48  &   4.62 \\ 
      Fe I   &  4073.763  &  3.27  &  $-$0.900  &   16.39  &   4.56 \\ 
      Fe I   &  4098.176  &  3.24  &  $-$0.880  &   18.97  &   4.58 \\ 
      Fe I   &  4114.445  &  2.83  &  $-$1.303  &   20.31  &   4.54 \\ 
      Fe I   &  4120.207  &  2.99  &  $-$1.270  &   16.12  &   4.58 \\ 
      Fe I   &  4121.802  &  2.83  &  $-$1.450  &   15.30  &   4.54 \\ 
      Fe I   &  4132.899  &  2.85  &  $-$1.010  &   24.62  &   4.38 \\ 
      Fe I   &  4134.678  &  2.83  &  $-$0.649  &   45.52  &   4.44 \\ 
      Fe I   &  4136.998  &  3.42  &  $-$0.450  &   24.65  &   4.51 \\ 
      Fe I   &  4139.927  &  0.99  &  $-$3.629  &   20.16  &   4.63 \\ 
      Fe I   &  4143.414  &  3.05  &  $-$0.200  &   52.61  &   4.40 \\ 
      Fe I   &  4143.868  &  1.56  &  $-$0.511  &  123.11  &   4.59 \\ 
      Fe I   &  4153.899  &  3.40  &  $-$0.320  &   26.19  &   4.39 \\ 
      Fe I   &  4154.498  &  2.83  &  $-$0.688  &   42.67  &   4.42 \\ 
      Fe I   &  4154.805  &  3.37  &  $-$0.400  &   25.64  &   4.42 \\ 
      Fe I   &  4156.799  &  2.83  &  $-$0.808  &   35.35  &   4.39 \\ 
      Fe I   &  4157.780  &  3.42  &  $-$0.403  &   23.58  &   4.42 \\ 
      Fe I   &  4158.793  &  3.43  &  $-$0.670  &   17.00  &   4.52 \\ 
      Fe I   &  4174.913  &  0.91  &  $-$2.938  &   54.65  &   4.55 \\ 
      Fe I   &  4181.755  &  2.83  &  $-$0.371  &   62.65  &   4.50 \\ 
      Fe I   &  4182.382  &  3.02  &  $-$1.180  &   15.52  &   4.49 \\ 
      Fe I   &  4184.892  &  2.83  &  $-$0.869  &   42.12  &   4.58 \\ 
      Fe I   &  4187.039  &  2.45  &  $-$0.514  &   81.84  &   4.61 \\ 
      Fe I   &  4187.795  &  2.42  &  $-$0.510  &   81.71  &   4.57 \\ 
      Fe I   &  4191.430  &  2.47  &  $-$0.666  &   65.71  &   4.42 \\ 
      Fe I   &  4195.329  &  3.33  &  $-$0.492  &   24.83  &   4.43 \\ 
      Fe I   &  4199.095  &  3.05  &   0.156  &   74.15  &   4.50 \\ 
      Fe I   &  4202.029  &  1.49  &  $-$0.689  &  115.75  &   4.48 \\ 
      Fe I   &  4217.545  &  3.43  &  $-$0.484  &   22.96  &   4.49 \\ 
      Fe I   &  4222.213  &  2.45  &  $-$0.914  &   59.17  &   4.48 \\ 
      Fe I   &  4227.427  &  3.33  &   0.266  &   65.76  &   4.52 \\ 
      Fe I   &  4233.603  &  2.48  &  $-$0.579  &   73.99  &   4.50 \\ 
      Fe I   &  4238.810  &  3.40  &  $-$0.233  &   35.90  &   4.49 \\ 
      Fe I   &  4250.119  &  2.47  &  $-$0.380  &   83.66  &   4.51 \\ 
      Fe I   &  4250.787  &  1.56  &  $-$0.713  &  119.93  &   4.65 \\ 
      Fe I   &  4260.474  &  2.40  &   0.077  &  109.20  &   4.56 \\ 
      Fe I   &  4271.154  &  2.45  &  $-$0.337  &   91.65  &   4.62 \\ 
      Fe I   &  4271.760  &  1.49  &  $-$0.173  &  145.53  &   4.51 \\ 
      Fe I   &  4325.762  &  1.61  &   0.006  &  149.85  &   4.52 \\ 
      Fe I   &  4337.046  &  1.56  &  $-$1.695  &   75.15  &   4.48 \\ 
      Fe I   &  4352.735  &  2.22  &  $-$1.290  &   56.59  &   4.49 \\ 
      Fe I   &  4375.930  &  0.00  &  $-$3.005  &  109.99  &   4.70 \\ 
      Fe I   &  4383.545  &  1.48  &   0.200  &  177.72  &   4.49 \\ 
      Fe I   &  4388.407  &  3.60  &  $-$0.681  &   11.49  &   4.49 \\ 
      Fe I   &  4404.750  &  1.56  &  $-$0.147  &  155.74  &   4.64 \\ 
      Fe I   &  4407.709  &  2.18  &  $-$1.970  &   34.42  &   4.68 \\ 
      Fe I   &  4415.122  &  1.61  &  $-$0.621  &  121.77  &   4.55 \\ 
      Fe I   &  4427.310  &  0.05  &  $-$2.924  &  109.71  &   4.65 \\ 
      Fe I   &  4430.614  &  2.22  &  $-$1.659  &   40.74  &   4.54 \\ 
      Fe I   &  4442.339  &  2.20  &  $-$1.228  &   67.20  &   4.58 \\ 
      Fe I   &  4443.194  &  2.86  &  $-$1.043  &   27.12  &   4.42 \\ 
      Fe I   &  4447.717  &  2.22  &  $-$1.339  &   55.97  &   4.50 \\ 
      Fe I   &  4454.381  &  2.83  &  $-$1.300  &   25.98  &   4.62 \\ 
      Fe I   &  4459.118  &  2.18  &  $-$1.279  &   66.05  &   4.58 \\ 
      Fe I   &  4461.653  &  0.09  &  $-$3.194  &  100.87  &   4.71 \\ 
      Fe I   &  4476.019  &  2.85  &  $-$0.820  &   46.66  &   4.57 \\ 
      Fe I   &  4484.220  &  3.60  &  $-$0.860  &   10.23  &   4.60 \\ 
      Fe I   &  4489.739  &  0.12  &  $-$3.899  &   66.33  &   4.65 \\ 
      Fe I   &  4494.563  &  2.20  &  $-$1.143  &   70.78  &   4.55 \\ 
      Fe I   &  4531.148  &  1.48  &  $-$2.101  &   69.10  &   4.58 \\ 
      Fe I   &  4592.651  &  1.56  &  $-$2.462  &   52.22  &   4.71 \\ 
      Fe I   &  4602.941  &  1.49  &  $-$2.208  &   66.23  &   4.63 \\ 
      Fe I   &  4630.120  &  2.28  &  $-$2.587  &    8.70  &   4.65 \\ 
      Fe I   &  4632.912  &  1.61  &  $-$2.913  &   23.49  &   4.66 \\ 
      Fe I   &  4647.434  &  2.95  &  $-$1.351  &   20.30  &   4.64 \\ 
      Fe I   &  4678.846  &  3.60  &  $-$0.830  &   12.03  &   4.62 \\ 
      Fe I   &  4691.411  &  2.99  &  $-$1.520  &   13.24  &   4.63 \\ 
      Fe I   &  4707.274  &  3.24  &  $-$1.080  &   17.19  &   4.62 \\ 
      Fe I   &  4710.283  &  3.02  &  $-$1.610  &    9.42  &   4.59 \\ 
      Fe I   &  4733.591  &  1.49  &  $-$2.988  &   26.27  &   4.64 \\ 
      Fe I   &  4736.772  &  3.21  &  $-$0.752  &   31.04  &   4.59 \\ 
      Fe I   &  4786.806  &  3.00  &  $-$1.606  &   10.75  &   4.62 \\ 
      Fe I   &  4859.741  &  2.88  &  $-$0.760  &   53.00  &   4.58 \\ 
      Fe I   &  4871.318  &  2.87  &  $-$0.362  &   72.14  &   4.52 \\ 
      Fe I   &  4872.137  &  2.88  &  $-$0.567  &   54.38  &   4.42 \\ 
      Fe I   &  4890.755  &  2.88  &  $-$0.394  &   72.42  &   4.56 \\ 
      Fe I   &  4891.492  &  2.85  &  $-$0.111  &   85.16  &   4.50 \\ 
      Fe I   &  4903.310  &  2.88  &  $-$0.926  &   37.13  &   4.46 \\ 
      Fe I   &  4918.994  &  2.85  &  $-$0.342  &   73.33  &   4.48 \\ 
      Fe I   &  4924.770  &  2.28  &  $-$2.114  &   24.33  &   4.65 \\ 
      Fe I   &  4938.814  &  2.88  &  $-$1.077  &   30.63  &   4.48 \\ 
      Fe I   &  4939.687  &  0.86  &  $-$3.252  &   57.14  &   4.65 \\ 
      Fe I   &  4946.388  &  3.37  &  $-$1.170  &   11.59  &   4.64 \\ 
      Fe I   &  4966.089  &  3.33  &  $-$0.871  &   20.26  &   4.58 \\ 
      Fe I   &  4994.130  &  0.92  &  $-$2.969  &   69.41  &   4.65 \\ 
      Fe I   &  5001.870  &  3.88  &   0.050  &   30.48  &   4.55 \\ 
      Fe I   &  5012.068  &  0.86  &  $-$2.642  &   96.10  &   4.79 \\ 
      Fe I   &  5041.072  &  0.96  &  $-$3.090  &   60.04  &   4.64 \\ 
      Fe I   &  5041.756  &  1.49  &  $-$2.200  &   75.76  &   4.70 \\ 
      Fe I   &  5049.820  &  2.28  &  $-$1.355  &   59.37  &   4.52 \\ 
      Fe I   &  5051.634  &  0.92  &  $-$2.764  &   85.91  &   4.75 \\ 
      Fe I   &  5068.766  &  2.94  &  $-$1.041  &   29.33  &   4.47 \\ 
      Fe I   &  5074.749  &  4.22  &  $-$0.200  &   10.07  &   4.60 \\ 
      Fe I   &  5079.224  &  2.20  &  $-$2.105  &   24.55  &   4.53 \\ 
      Fe I   &  5079.740  &  0.99  &  $-$3.245  &   57.51  &   4.79 \\ 
      Fe I   &  5083.339  &  0.96  &  $-$2.842  &   73.53  &   4.63 \\ 
      Fe I   &  5098.697  &  2.18  &  $-$2.030  &   43.97  &   4.81 \\ 
      Fe I   &  5110.413  &  0.00  &  $-$3.760  &   95.98  &   4.79 \\ 
      Fe I   &  5127.360  &  0.92  &  $-$3.249  &   56.69  &   4.68 \\ 
      Fe I   &  5131.468  &  2.22  &  $-$2.515  &   12.58  &   4.62 \\ 
      Fe I   &  5133.689  &  4.18  &   0.140  &   18.84  &   4.52 \\ 
      Fe I   &  5142.929  &  0.96  &  $-$3.080  &   66.14  &   4.73 \\ 
      Fe I   &  5150.839  &  0.99  &  $-$3.037  &   59.45  &   4.60 \\ 
      Fe I   &  5151.911  &  1.01  &  $-$3.321  &   51.73  &   4.78 \\ 
      Fe I   &  5166.282  &  0.00  &  $-$4.123  &   78.79  &   4.79 \\ 
      Fe I   &  5171.596  &  1.49  &  $-$1.721  &   96.70  &   4.62 \\ 
      Fe I   &  5191.455  &  3.04  &  $-$0.551  &   56.08  &   4.57 \\ 
      Fe I   &  5192.344  &  3.00  &  $-$0.421  &   56.61  &   4.40 \\ 
      Fe I   &  5194.942  &  1.56  &  $-$2.021  &   83.02  &   4.72 \\ 
      Fe I   &  5198.711  &  2.22  &  $-$2.091  &   24.66  &   4.53 \\ 
      Fe I   &  5202.336  &  2.18  &  $-$1.871  &   42.80  &   4.62 \\ 
      Fe I   &  5216.274  &  1.61  &  $-$2.082  &   66.28  &   4.52 \\ 
      Fe I   &  5225.526  &  0.11  &  $-$4.755  &   33.12  &   4.76 \\ 
      Fe I   &  5232.940  &  2.94  &  $-$0.057  &   82.71  &   4.43 \\ 
      Fe I   &  5242.491  &  3.63  &  $-$0.967  &    8.44  &   4.56 \\ 
      Fe I   &  5247.050  &  0.09  &  $-$4.946  &   28.26  &   4.84 \\ 
      Fe I   &  5250.210  &  0.12  &  $-$4.938  &   21.48  &   4.71 \\ 
      Fe I   &  5250.646  &  2.20  &  $-$2.180  &   34.67  &   4.80 \\ 
      Fe I   &  5254.956  &  0.11  &  $-$4.764  &   37.38  &   4.85 \\ 
      Fe I   &  5266.555  &  3.00  &  $-$0.385  &   65.34  &   4.51 \\ 
      Fe I   &  5269.537  &  0.86  &  $-$1.333  &  159.43  &   4.76 \\ 
      Fe I   &  5281.790  &  3.04  &  $-$0.833  &   41.05  &   4.58 \\ 
      Fe I   &  5283.621  &  3.24  &  $-$0.524  &   41.32  &   4.52 \\ 
      Fe I   &  5302.300  &  3.28  &  $-$0.720  &   30.24  &   4.56 \\ 
      Fe I   &  5307.361  &  1.61  &  $-$2.912  &   25.00  &   4.60 \\ 
      Fe I   &  5324.179  &  3.21  &  $-$0.103  &   61.72  &   4.40 \\ 
      Fe I   &  5328.039  &  0.92  &  $-$1.466  &  152.54  &   4.82 \\ 
      Fe I   &  5328.531  &  1.56  &  $-$1.850  &   92.61  &   4.71 \\ 
      Fe I   &  5332.900  &  1.55  &  $-$2.776  &   31.82  &   4.54 \\ 
      Fe I   &  5339.930  &  3.27  &  $-$0.720  &   30.56  &   4.55 \\ 
      Fe I   &  5364.871  &  4.45  &     0.228  &   12.87  &   4.54 \\ 
      Fe I   &  5365.400  &  3.56  &  $-$1.020  &   11.05  &   4.65 \\ 
      Fe I   &  5371.489  &  0.96  &  $-$1.644  &  148.24  &   4.95 \\ 
      Fe I   &  5383.369  &  4.31  &     0.645  &   35.53  &   4.52 \\ 
      Fe I   &  5393.168  &  3.24  &  $-$0.910  &   32.33  &   4.73 \\ 
      Fe I   &  5397.128  &  0.92  &  $-$1.982  &  129.30  &   4.83 \\ 
      Fe I   &  5405.775  &  0.99  &  $-$1.852  &  136.75  &   4.95 \\ 
      Fe I   &  5415.199  &  4.39  &     0.643  &   35.76  &   4.63 \\ 
      Fe I   &  5424.068  &  4.32  &     0.520  &   33.18  &   4.61 \\ 
      Fe I   &  5429.696  &  0.96  &  $-$1.881  &  132.11  &   4.84 \\ 
      Fe I   &  5434.524  &  1.01  &  $-$2.126  &  116.05  &   4.79 \\ 
      Fe I   &  5446.917  &  0.99  &  $-$1.910  &  132.28  &   4.90 \\ 
      Fe I   &  5455.609  &  1.01  &  $-$2.090  &  123.53  &   4.90 \\ 
      Fe I   &  5497.516  &  1.01  &  $-$2.825  &   80.46  &   4.74 \\ 
      Fe I   &  5501.465  &  0.96  &  $-$3.046  &   75.60  &   4.80 \\ 
      Fe I   &  5506.779  &  0.99  &  $-$2.789  &   81.84  &   4.69 \\ 
      Fe I   &  5569.618  &  3.42  &  $-$0.540  &   29.34  &   4.50 \\ 
      Fe I   &  5572.842  &  3.40  &  $-$0.275  &   45.50  &   4.52 \\ 
      Fe I   &  5576.088  &  3.43  &  $-$1.000  &   14.19  &   4.58 \\ 
      Fe I   &  5586.756  &  3.37  &  $-$0.144  &   55.13  &   4.50 \\ 
      Fe I   &  5615.644  &  3.33  &     0.050  &   70.83  &   4.53 \\ 
      Fe I   &  5658.816  &  3.40  &  $-$0.793  &   22.66  &   4.58 \\ 
      Fe I   &  5662.516  &  4.18  &  $-$0.573  &    6.78  &   4.69 \\ 
      Fe I   &  5701.544  &  2.56  &  $-$2.143  &   12.34  &   4.59 \\ 
      Fe I   &  6065.481  &  2.61  &  $-$1.410  &   37.90  &   4.51 \\ 
      Fe I   &  6136.615  &  2.45  &  $-$1.410  &   52.64  &   4.55 \\ 
      Fe I   &  6137.691  &  2.59  &  $-$1.346  &   41.54  &   4.49 \\ 
      Fe I   &  6191.558  &  2.43  &  $-$1.416  &   56.39  &   4.59 \\ 
      Fe I   &  6219.280  &  2.20  &  $-$2.448  &   21.40  &   4.70 \\ 
      Fe I   &  6230.723  &  2.56  &  $-$1.276  &   51.97  &   4.54 \\ 
      Fe I   &  6252.555  &  2.40  &  $-$1.687  &   42.20  &   4.59 \\ 
      Fe I   &  6254.257  &  2.28  &  $-$2.443  &   24.66  &   4.86 \\ 
      Fe I   &  6265.134  &  2.18  &  $-$2.540  &   24.31  &   4.83 \\ 
      Fe I   &  6335.330  &  2.20  &  $-$2.180  &   32.04  &   4.65 \\ 
      Fe I   &  6393.601  &  2.43  &  $-$1.576  &   46.87  &   4.59 \\ 
      Fe I   &  6411.649  &  3.65  &  $-$0.595  &   21.03  &   4.58 \\ 
      Fe I   &  6421.350  &  2.28  &  $-$2.014  &   33.60  &   4.60 \\ 
      Fe I   &  6430.846  &  2.18  &  $-$1.946  &   43.03  &   4.58 \\ 
      Fe I   &  6494.980  &  2.40  &  $-$1.239  &   76.18  &   4.67 \\ 
      Fe I   &  6592.912  &  2.73  &  $-$1.473  &   30.35  &   4.54 \\ 
      Fe I   &  6663.440  &  2.42  &  $-$2.479  &   11.55  &   4.66 \\ 
      Fe I   &  6677.986  &  2.69  &  $-$1.418  &   37.18  &   4.56 \\ 
      Fe I   &  6978.850  &  2.48  &  $-$2.452  &   10.98  &   4.66 \\ 
     Fe II   &  4489.185  &  2.83  &  $-$2.970  &   14.64  &   4.59 \\ 
     Fe II   &  4491.410  &  2.86  &  $-$2.710  &   19.12  &   4.51 \\ 
     Fe II   &  4515.340  &  2.84  &  $-$2.600  &   28.14  &   4.59 \\ 
     Fe II   &  4520.224  &  2.81  &  $-$2.600  &   31.25  &   4.62 \\ 
     Fe II   &  4541.523  &  2.86  &  $-$3.050  &   10.75  &   4.54 \\ 
     Fe II   &  4555.890  &  2.83  &  $-$2.400  &   35.36  &   4.52 \\ 
     Fe II   &  4576.340  &  2.84  &  $-$2.950  &   16.37  &   4.63 \\ 
     Fe II   &  4583.840  &  2.81  &  $-$1.930  &   60.14  &   4.47 \\ 
     Fe II   &  4923.930  &  2.89  &  $-$1.320  &   93.25  &   4.53 \\ 
     Fe II   &  5197.580  &  3.23  &  $-$2.220  &   26.54  &   4.56 \\ 
     Fe II   &  5234.630  &  3.22  &  $-$2.180  &   32.27  &   4.62 \\ 
     Fe II   &  5276.000  &  3.20  &  $-$2.010  &   42.66  &   4.61 \\ 
     Fe II   &  5284.080  &  2.89  &  $-$3.190  &   10.78  &   4.64 \\ 
      Co I   &  3842.047  &  0.92  &  $-$0.770  &   47.49  &   2.23 \\ 
      Co I   &  3845.468  &  0.92  &   0.010  &   81.12  &   2.30 \\ 
      Co I   &  3881.869  &  0.58  &  $-$1.130  &   52.85  &   2.27 \\ 
      Co I   &  3995.306  &  0.92  &  $-$0.220  &   73.10  &   2.19 \\ 
      Co I   &  4118.767  &  1.05  &  $-$0.490  &   60.17  &   2.26 \\ 
      Co I   &  4121.318  &  0.92  &  $-$0.320  &   74.45  &   2.25 \\ 
      Ni I   &  3500.850  &  0.17  &  $-$1.294  &   88.63  &   3.30 \\ 
      Ni I   &  3524.540  &  0.03  &   0.007  &  171.36  &   3.21 \\ 
      Ni I   &  3566.370  &  0.42  &  $-$0.251  &  116.06  &   3.20 \\ 
      Ni I   &  3597.710  &  0.21  &  $-$1.115  &   99.47  &   3.31 \\ 
      Ni I   &  3807.140  &  0.42  &  $-$1.220  &   97.22  &   3.29 \\ 
      Ni I   &  3858.301  &  0.42  &  $-$0.951  &  108.26  &   3.25 \\ 
      Ni I   &  4648.659  &  3.42  &  $-$0.160  &    8.30  &   3.27 \\ 
      Ni I   &  4714.421  &  3.38  &   0.230  &   17.96  &   3.21 \\ 
      Ni I   &  4904.410  &  3.54  &  $-$0.170  &    6.04  &   3.24 \\ 
      Ni I   &  4980.161  &  3.61  &  $-$0.110  &    6.33  &   3.28 \\ 
      Ni I   &  5080.523  &  3.65  &   0.130  &    7.95  &   3.19 \\ 
      Ni I   &  5084.080  &  3.68  &   0.030  &    5.35  &   3.14 \\ 
      Ni I   &  5476.900  &  1.83  &  $-$0.890  &   60.32  &   3.18 \\ 
      Ni I   &  5754.675  &  1.94  &  $-$2.330  &    3.10  &   3.09 \\ 
      Ni I   &  6108.121  &  1.68  &  $-$2.450  &    5.53  &   3.12 \\ 
      Zn I   &  4722.150  &  4.03  &  $-$0.390  &    7.68  &   1.75 \\ 
      Zn I   &  4810.528  &  4.08  &  $-$0.137  &   10.51  &   1.70 \\ 
\enddata
\end{deluxetable}

\startlongtable
\begin{deluxetable}{lrrrr}
\tabletypesize{\scriptsize}
\tablewidth{0pc}
\tablecaption{\label{tab:ncap} Individual Abundance Measurements of
Neutron-Capture Elements for \protect\rii}
\tablehead{
\colhead{Ion}&
\colhead{$\lambda$}&
\colhead{$\chi$} &
\colhead{$\log\,gf$}&
\colhead{$\log\epsilon$\,(X)}\\
\colhead{}&
\colhead{({\AA})}&
\colhead{(eV)} &
\colhead{}&
\colhead{}}
\startdata
\ion{Sr}{2}	&	4161.792	&	2.94	&	$-$0.502	&	0.56  \\
\ion{Sr}{2}	&	4215.519	&	0.00	&	$-$0.170	&	0.43  \\
\ion{Y}{2}	&	3747.556	&	0.10	&	$-$0.910	&	$-$0.34 \\
\ion{Y}{2}	&	4398.013	&	0.13	&	$-$1.000	&	$-$0.49 \\
\ion{Y}{2}	&	4682.324	&	0.41	&	$-$1.510	&	$-$0.38 \\
\ion{Y}{2}	&	4883.684	&	1.08	&	0.070	&	$-$0.52 \\
\ion{Y}{2}	&	4900.120	&	1.03	&	$-$0.090	&	$-$0.72 \\
\ion{Zr}{2}	&	3573.055	&	0.32	&	$-$1.041	&	0.42 \\
\ion{Zr}{2}	&	3836.761	&	0.56	&	$-$0.120	&	0.38 \\
\ion{Zr}{2}	&	3991.127	&	0.76	&	$-$0.310	&	0.28 \\
\ion{Zr}{2}	&	3998.954	&	0.56	&	$-$0.520	&	0.36 \\
\ion{Zr}{2}	&	4317.299	&	0.71	&	$-$1.450	&	0.38 \\
\ion{Zr}{2}	&	4050.316	&	0.71	&	$-$1.060	&	0.38 \\
\ion{Ba}{2}	&	5853.675	&	0.60	&	$-$1.010	&	$-$0.03\\
\ion{Ba}{2}	&	6141.713	&	0.70	&	$-$0.077	&	0.08\\
\ion{Ba}{2}	&	6496.898	&	0.60	&	$-$0.380	&	0.25\\
\ion{La}{2}	&	3794.774	&	0.24	&	0.140	&	$-$0.60\\
\ion{La}{2}	&	3988.515	&	0.40	&	0.170	&	$-$0.80\\
\ion{La}{2}	&	3995.745	&	0.17	&	$-$0.100	&	$-$0.70\\
\ion{La}{2}	&	4086.709	&	0.00	&	0.230	&	$-$0.70\\
\ion{La}{2}	&	4123.218	&	0.32	&	0.110	&	$-$0.85\\
\ion{La}{2}	&	4429.905	&	0.24	&	$-$1.490	&	$-$0.90\\
\ion{Ce}{2}	&	3940.330	&	0.32	&	$-$0.270	&	$-$0.35\\
\ion{Ce}{2}	&	3942.151	&	0.000	&	$-$0.22	&	$-$0.37\\
\ion{Ce}{2}	&	3999.237	&	0.30	&	0.090	&	$-$0.23\\
\ion{Ce}{2}	&	4014.897	&	0.53	&	0.140	&	$-$0.48\\
\ion{Ce}{2}	&	4072.918	&	0.33	&	$-$0.710	&	$-$0.14\\
\ion{Ce}{2}	&	4073.474	&	0.48	&	0.230	&	$-$0.32\\
\ion{Ce}{2}	&	4137.645	&	0.52	&	0.440	&	$-$0.35\\
\ion{Ce}{2}	&	4138.096	&	0.924	&	$-$0.08	&	$-$0.32\\
\ion{Ce}{2}	&	4165.599	&	0.91	&	1.420	&	$-$0.36\\
\ion{Ce}{2}	&	4222.597	&	0.12	&	0.020	&	$-$0.41\\
\ion{Ce}{2}	&	4562.359	&	0.48	&	0.381	&	$-$0.37\\
\ion{Ce}{2}	&	4449.330	&	0.61	&	0.080	&	$-$0.51\\
\ion{Pr}{2}	&	3964.812	&	0.06	&	$-$0.180	&	$-$0.93 \\
\ion{Pr}{2}	&	3965.253	&	0.20	&	$-$0.195	&	$-$0.88 \\
\ion{Pr}{2}	&	4179.393	&	0.20	&	0.293	&	$-$0.88 \\
\ion{Pr}{2}	&	4189.479	&	0.37	&	0.175	&	$-$0.86 \\
\ion{Pr}{2}	&	4222.934	&	0.06	&	0.018	&	$-$0.88 \\
\ion{Pr}{2}	&	4408.819	&	0.00	&	$-$0.278	&	$-$0.92 \\
\ion{Pr}{2}	&	4449.823	&	0.20	&	$-$0.436	&	$-$0.90 \\
\ion{Nd}{2}	&	3862.566	&	0.18	&	$-$0.760	&	$-$0.18\\
\ion{Nd}{2}	&	3863.408	&	0.00	&	$-$0.010	&	$-$0.23\\
\ion{Nd}{2}	&	3900.215	&	0.47	&	0.100	&	$-$0.24\\
\ion{Nd}{2}	&	3991.735	&	0.00	&	$-$0.260	&	$-$0.14\\
\ion{Nd}{2}	&	4021.728	&	0.18	&	$-$0.310	&	$-$0.37\\
\ion{Nd}{2}	&	4051.139	&	0.38	&	$-$0.300	&	$-$0.12\\
\ion{Nd}{2}	&	4061.080	&	0.47	&	1.380	&	$-$0.12\\
\ion{Nd}{2}	&	4110.470	&	0.00	&	$-$0.710	&	$-$0.18\\
\ion{Nd}{2}	&	4179.580	&	0.18	&	$-$0.640	&	$-$0.23\\
\ion{Nd}{2}	&	4178.635	&	0.18	&	$-$1.030	&	$-$0.14\\
\ion{Nd}{2}	&	4177.320	&	0.06	&	$-$0.100	&	$-$0.12\\
\ion{Sm}{2}	&	3896.970	&	0.04	&	$-$0.670	&	$-$0.56 \\
\ion{Sm}{2}	&	4188.128	&	0.54	&	$-$0.440	&	$-$0.40 \\
\ion{Sm}{2}	&	4318.926	&	0.28	&	$-$0.250	&	$-$0.55 \\
\ion{Sm}{2}	&	4424.337	&	0.49	&	0.140	&	$-$0.58 \\
\ion{Sm}{2}	&	4421.126	&	0.38	&	$-$0.490	&	$-$0.58 \\
\ion{Eu}{2}	&	3724.930	&	0.00	&	$-$0.090	&	$-$0.74 \\
\ion{Eu}{2}	&	3907.107	&	0.21	&	0.170	&	$-$0.78 \\
\ion{Eu}{2}	&	4129.720	&	0.00	&	0.220	&	$-$0.66 \\
\ion{Eu}{2}	&	4205.040	&	0.00	&	0.210	&	$-$0.73 \\
\ion{Eu}{2}	&	4435.578	&	0.21	&	$-$0.110	&	$-$0.86 \\
\ion{Eu}{2}	&	6645.060	&	1.38	&	0.120	&	$-$0.78 \\
\ion{Gd}{2}	&	3549.359	&	0.24	&	0.290	&	$-$0.39 \\
\ion{Gd}{2}	&	3697.733	&	0.03	&	$-$0.340	&	$-$0.40 \\
\ion{Gd}{2}	&	3768.396	&	0.08	&	0.210	&	$-$0.35 \\
\ion{Gd}{2}	&	3796.384	&	0.03	&	0.020	&	$-$0.29 \\
\ion{Gd}{2}	&	3844.578	&	0.14	&	$-$0.460	&	$-$0.33 \\
\ion{Gd}{2}	&	4191.075	&	0.43	&	$-$0.480	&	$-$0.29 \\
\ion{Gd}{2}	&	4215.022	&	0.43	&	$-$0.440	&	$-$0.39 \\
\ion{Gd}{2}	&	4251.731	&	0.38	&	$-$0.220	&	$-$0.49 \\
\ion{Tb}{2}	&	3702.850	&	0.13	&	0.440	&	$-$1.30 \\
\ion{Tb}{2}	&	3747.380	&	0.40	&	0.130	&	$-$1.30 \\
\ion{Tb}{2}	&	3848.730	&	0.00	&	0.280	&	$-$1.20 \\
\ion{Tb}{2}	&	3874.168	&	0.00	&	0.270	&	$-$1.26 \\
\ion{Tb}{2}	&	4002.566	&	0.64	&	0.100	&	$-$1.08 \\
\ion{Dy}{2}	&	3757.368	&	0.10	&	$-$0.170	&	$-$0.20 \\
\ion{Dy}{2}	&	3944.680	&	0.00	&	0.110	&	$-$0.24 \\
\ion{Dy}{2}	&	3996.689	&	0.59	&	$-$0.260	&	$-$0.42 \\
\ion{Dy}{2}	&	4050.565	&	0.59	&	$-$0.470	&	$-$0.40 \\
\ion{Dy}{2}	&	4073.120	&	0.54	&	$-$0.320	&	$-$0.44 \\
\ion{Dy}{2}	&	4077.966	&	0.10	&	$-$0.040	&	$-$0.26 \\
\ion{Dy}{2}	&	4103.306	&	0.10	&	$-$0.380	&	$-$0.26 \\
\ion{Dy}{2}	&	4449.700	&	0.00	&	$-$1.030	&	$-$0.42 \\
\ion{Ho}{2}	&	3796.730	&	0.00	&	0.160	&	$-$1.02 \\
\ion{Ho}{2}	&	3810.738	&	0.00	&	0.142	&	$-$1.00 \\
\ion{Ho}{2}	&	3890.970	&	0.08	&	0.460	&	$-$1.06 \\
\ion{Er}{2}	&	3692.649	&	0.06	&	0.138	&	$-$0.28 \\
\ion{Er}{2}	&	3906.311	&	0.00	&	$-$0.052	&	$-$0.28 \\
\ion{Er}{2}	&	3729.524	&	0.00	&	$-$0.488	&	$-$0.47 \\
\ion{Er}{2}	&	3786.836	&	0.00	&	$-$0.644	&	$-$0.35 \\
\ion{Er}{2}	&	3830.481	&	0.00	&	$-$0.365	&	$-$0.47 \\
\ion{Er}{2}	&	3896.233	&	0.06	&	$-$0.241	&	$-$0.48 \\
\ion{Er}{2}	&	3938.626	&	0.00	&	$-$0.610	&	$-$0.43 \\
\ion{Tm}{2}	&	3700.255	&	0.03	&	$-$0.380	&	$-$1.19 \\
\ion{Tm}{2}	&	3701.362	&	0.00	&	$-$0.540	&	$-$1.20 \\
\ion{Tm}{2}	&	3795.759	&	0.03	&	$-$0.230	&	$-$1.21 \\
\ion{Tm}{2}	&	3848.019	&	0.00	&	$-$0.140	&	$-$1.26 \\
\ion{Tm}{2}	&	3996.510	&	0.00	&	$-$1.200	&	$-$1.20 \\
\ion{Yb}{2}	&	3694.190	&	0.00	&	$-$0.320	&	$-$0.51 \\
\ion{Lu}{2}	&	3472.476	&	1.54	&	$-$0.220	&	$-$1.35 \\
\ion{Lu}{2}	&	3507.380	&	0.00	&	$-$1.160	&	$-$1.50 \\
\ion{Hf}{2}	&	3719.276	&	0.61	&	$-$0.810	&	$-$0.50 \\
\ion{Hf}{2}	&	3793.379	&	0.37	&	$-$1.110	&	$-$0.68 \\
\ion{Hf}{2}	&	3918.090	&	0.45	&	$-$1.140	&	$-$0.69 \\
\ion{Hf}{2}	&	4093.150	&	0.45	&	$-$1.150	&	$-$0.79 \\
\ion{Os}{1}	&	4135.775	&	0.52	&	$-$1.260	&	0.31 \\
\ion{Os}{1}	&	4260.848	&	0.00	&	$-$1.440	&	$-$0.14 \\
\ion{Os}{1}	&	4420.520	&	0.33	&	$-$0.430	&	0.11 \\
\ion{Ir}{1}	&	3800.120	&	0.00	&	$-$1.450	&	$-$0.02 \\
\ion{Pb}{1}	&	4057.807	&	1.32	&	$-$0.170	&	0.12 \\
\ion{Th}{2}	&	4094.747	&	0.00	&	$-$0.885	&	$-$1.17\\
\ion{Th}{2}	&	4086.521	&	0.00	&	$-$0.929	&	$-$1.28\\
\ion{Th}{2}	&	4019.129	&	0.00	&	$-$0.228	&	$-$1.27\\
\ion{U}{2}	&	3859.571	&	0.036	&	$-$0.07	&	$-$2.14\\
\enddata
\end{deluxetable}

\clearpage

\startlongtable
\begin{deluxetable}{@{}l@{}rrrrrrr@{}}  
\tabletypesize{\small}
\tablecaption{Final Abundance Estimates for \protect\rii \label{abfinal}} 
\tablehead{Ion & $\log\epsilon_{\odot}$\,(X)  & $\log\epsilon$\,(X) 
               & $\mbox{[X/H]}$ & $\mbox{[X/Fe]}$ & $\sigma$ &
			   $\overline{\sigma}$ & $n$}
\startdata 
C (CH)      &   8.43    &   5.08    &   $-$3.35 &  $-$0.44    &  0.20 & 0.20   &    1 \\
C (CH)      &   8.43    &   5.75    &   $-$2.68 &  $+$0.23\xx &  0.20 & 0.20   &    1 \\
\ion{Na}{1} &   6.24    &   3.72    &   $-$2.52 &  $+$0.39  &    0.10 & 0.10   &    1 \\
\ion{Mg}{1} &   7.60    &   5.05    &   $-$2.55 &  $+$0.36  &    0.05 & 0.10   &   10 \\
\ion{Al}{1} &   6.45    &   3.09    &   $-$3.36 &  $-$0.45  &    0.10 & 0.10   &    1 \\
\ion{Si}{1} &   7.51    &   5.19    &   $-$2.32 &  $+$0.59  &    0.03 & 0.10   &    2 \\
\ion{Ca}{1} &   6.34    &   3.59    &   $-$2.75 &  $+$0.16  &    0.07 & 0.10   &   13 \\
\ion{Sc}{2} &   3.15    &   0.21    &   $-$2.94 &  $-$0.03  &    0.04 & 0.10   &    8 \\
\ion{Ti}{1} &   4.95    &   2.12    &   $-$2.83 &  $+$0.08  &    0.03 & 0.10   &   12 \\
\ion{Ti}{2} &   4.95    &   2.16    &   $-$2.79 &  $+$0.12  &    0.03 & 0.10   &   34 \\
\ion{V}{2}  &   3.93    &   1.30    &   $-$2.63 &  $+$0.28  &    0.04 & 0.10   &    2 \\
\ion{Cr}{1} &   5.64    &   2.45    &   $-$3.19 &  $-$0.28  &    0.04 & 0.10   &   16 \\
\ion{Mn}{1} &   5.43    &   2.01    &   $-$3.42 &  $-$0.51  &    0.02 & 0.10   &    7 \\
\ion{Fe}{1} &   7.50    &   4.59    &   $-$2.91 &  $+$0.00  &    0.12 & 0.10   &  202 \\
\ion{Fe}{2} &   7.50    &   4.57    &   $-$2.93 &  $-$0.02  &    0.05 & 0.10   &   13 \\
\ion{Co}{1} &   4.99    &   2.25    &   $-$2.74 &  $+$0.17  &    0.04 & 0.10   &    6 \\
\ion{Ni}{1} &   6.22    &   3.22    &   $-$3.00 &  $-$0.09  &    0.07 & 0.10   &   15 \\
\ion{Zn}{1} &   4.56    &   1.73    &   $-$2.83 &  $+$0.08  &    0.04 & 0.10   &    2 \\
\ion{Sr}{2}	&	2.87	&	0.50	&	$-$2.38	& $+$0.54	&	0.11 & 0.11	&	2 \\
\ion{Y}{2}	&	2.21	&	$-$0.49	&	$-$2.70	& $+$0.21	&	0.07 & 0.10	&	5 \\
\ion{Zr}{2}	&	2.58	&	0.37	&	$-$2.21	& $+$0.70	&	0.02 & 0.10	&	6 \\
\ion{Ba}{2}	&	2.18	&	0.10	&	$-$2.08	& $+$0.83	&	0.10 & 0.10	&	3 \\
\ion{La}{2}	&	1.10	&	$-$0.76	&	$-$1.86	& $+$1.05	&	0.07 & 0.10	&   6 \\
\ion{Ce}{2}	&	1.58	&	$-$0.35	&	$-$1.93	& $+$0.98	&	0.04 & 0.10	&	12 \\
\ion{Pr}{2}	&	0.72	&	$-$0.89	&	$-$1.61	& $+$1.30	&	0.01 & 0.10	&	7 \\
\ion{Nd}{2}	&	1.42	&	$-$0.19	&	$-$1.61	& $+$1.30	&	0.03 & 0.10	&	11 \\
\ion{Sm}{2}	&	0.96	&	$-$0.53	&	$-$1.49	& $+$1.42	&	0.03 & 0.10	&	5 \\
\ion{Eu}{2}	&	0.52	&	$-$0.75	&	$-$1.27	& $+$1.64	&	0.04 & 0.10	&	6 \\
\ion{Gd}{2}	&	1.07	&	$-$0.37	&	$-$1.44	& $+$1.47	&	0.02 & 0.10	&	8 \\
\ion{Tb}{2}	&	0.30	&	$-$1.23	&	$-$1.53	& $+$1.38	&	0.04 & 0.10	&	5 \\
\ion{Dy}{2}	&	1.10	&	$-$0.33	&	$-$1.43	& $+$1.48	&	0.03 & 0.10	&	8 \\
\ion{Ho}{2}	&	0.48	&	$-$1.03	&	$-$1.51	& $+$1.40	&	0.02 & 0.10	&	3 \\
\ion{Er}{2}	&	0.92	&	$-$0.39	&	$-$1.31	& $+$1.60	&	0.03 & 0.10	&	7 \\
\ion{Tm}{2}	&	0.10	&	$-$1.21	&	$-$1.31	& $+$1.60	&	0.01 & 0.10	&	5 \\
\ion{Yb}{2}	&	0.84	&	$-$0.51	&	$-$1.35	& $+$1.56	&	0.20 & 0.20	&	1 \\
\ion{Lu}{2}	&	0.10	&	$-$1.43	&	$-$1.53	& $+$1.39	&	0.09 & 0.10	&	2 \\
\ion{Hf}{2}	&	0.85	&	$-$0.67	&	$-$1.52	& $+$1.40	&	0.07 & 0.10	&	4 \\
\ion{Os}{1}	&	1.40	&	0.09	&	$-$1.31	& $+$1.60	&	0.15 & 0.15	&	3 \\
\ion{Ir}{1}	&	1.38	&	$-$0.02	&	$-$1.40	& $+$1.51	&	0.20 & 0.20	&	1 \\
\ion{Pb}{1}	&	1.75	&	$-$0.10	&	$-$1.85	& $+$1.06	&	0.20 & 0.20	&	1 \\
\ion{Th}{2}	&	0.02	&	$-$1.24	&	$-$1.26	& $+$1.65	&	0.04 & 0.10	&	3 \\
\ion{U}{2}	&	$-$0.54	&	$-$2.14	&	$-$1.60	& $+$1.31	&	0.20 & 0.20	&	1 \\
\enddata
\tablenotetext{a}{\cfe=$+$0.23 using corrections of \citet{placco2014c}.}
\end{deluxetable}

\begin{deluxetable}{lrrrrr}
\tabletypesize{\scriptsize}
\tablewidth{0pc}
\tablecaption{Example Systematic Abundance Uncertainties for \protect\rii \label{sys}}
\tablehead{
\colhead{Elem}&
\colhead{$\Delta$\teff}&
\colhead{$\Delta$\logg}&
\colhead{$\Delta\xi$}&
\colhead{$\sigma/\sqrt{n}$}&
\colhead{$\sigma_{\rm tot}$}\\
\colhead{}&
\colhead{$+$100\,K}&
\colhead{$+$0.2 dex}&
\colhead{$+$0.2 km/s}&
\colhead{}&
\colhead{}}
\startdata
Na I   &    0.10   & $-$0.10   & $-$0.11   &    0.10   &    0.21 \\
 Mg I   &    0.08   & $-$0.08   & $-$0.04   &    0.04   &    0.13 \\
 Al I   &    0.03   & $-$0.16   & $-$0.11   &    0.10   &    0.22 \\
 Si I   &    0.07   & $-$0.06   & $-$0.04   &    0.10   &    0.14 \\
  K I   &    0.09   & $-$0.01   & $-$0.01   &    0.10   &    0.14 \\
 Ca I   &    0.07   & $-$0.03   & $-$0.02   &    0.03   &    0.08 \\
Sc II   &    0.03   &    0.03   & $-$0.03   &    0.04   &    0.07 \\
 Ti I   &    0.12   & $-$0.03   & $-$0.02   &    0.04   &    0.13 \\
Ti II   &    0.01   &    0.02   & $-$0.05   &    0.02   &    0.06 \\
 Cr I   &    0.11   & $-$0.04   & $-$0.04   &    0.03   &    0.13 \\
 Mn I   &    0.05   & $-$0.14   & $-$0.17   &    0.06   &    0.23 \\
 Fe I   &    0.10   & $-$0.05   & $-$0.05   &    0.01   &    0.12 \\
Fe II   & $-$0.01   &    0.04   & $-$0.01   &    0.04   &    0.06 \\
 Co I   &    0.09   & $-$0.07   & $-$0.06   &    0.05   &    0.14 \\
 Ni I   &    0.10   & $-$0.01   & $-$0.01   &    0.04   &    0.11 \\
 Zn I   &    0.03   &    0.02   & $-$0.00   &    0.10   &    0.11 \\
Sr II   &    0.06   & $-$0.05   & $-$0.12   &    0.10   &    0.17 \\
Ba II   &    0.09   & $-$0.03   & $-$0.11   &    0.06   &    0.16 \\
\enddata
\end{deluxetable}

\begin{deluxetable}{lrrrrrr}
\tabletypesize{\scriptsize}
\tablewidth{0pc}
\tablecaption{\label{tab:ages} Ages of \protect{\rii} Calculated from Th
and U Chronometer Pairs 
}
\tablehead{
\colhead{X/Y}&
\colhead{$\eps{X/Y}_{\text{obs}}$}&
\colhead{PR}&
\colhead{Age}&
\colhead{PR}&
\colhead{Age}&
\colhead{$\sigma$}\\
\colhead{}&
\colhead{}&
\colhead{(i)}&
\colhead{(Gyr)}&
\colhead{(ii)}&
\colhead{(Gyr)} &
\colhead{(Gyr)}
}
\startdata
Th/Ba\xx&	$-1.34\pm 0.10$	&	\nodata	&	\nodata	&	$-$1.058	&	13.16	&	4.80 \\
Th/La	&	$-0.48\pm 0.08$	&	$-$0.60	&	$-$5.52	&	$-$0.362	&	5.58	&	3.83 \\
Th/Ce	&	$-0.89\pm 0.05$	&	$-$0.79	&	4.63	&	$-$0.724	&	7.71	&	2.50 \\
Th/Pr	&	$-0.35\pm 0.04$	&	$-$0.30	&	2.20	&	$-$0.313	&	1.59	&	1.81 \\
Th/Nd	&	$-1.05\pm 0.05$	&	$-$0.91	&	6.62	&	$-$0.928	&	5.78	&	2.13 \\
Th/Sm	&	$-0.71\pm 0.05$	&	$-$0.61	&	4.48	&	$-$0.796	&	$-$4.20	&	2.38 \\
Th/Eu\xx&	$-0.49\pm 0.05$	&	$-$0.33	&	7.28	&	$-$0.240	&	11.48	&	2.51 \\
Th/Gd\xx&	$-0.87\pm 0.05$	&	$-$0.81	&	2.98	&	$-$0.569	&	14.22	&	2.10 \\
Th/Tb	&	$-0.01\pm 0.06$	&	$-$0.12	&	$-$5.04	&	\nodata	&	\nodata	&	2.64 \\
Th/Dy	&	$-0.91\pm 0.05$	&	$-$0.89	&	0.93	&	$-$0.827	&	3.87	&	2.24 \\
Th/Ho	&	$-0.21\pm 0.04$	&	\nodata	&	\nodata	&	$-$0.071	&	6.64	&	2.00 \\
Th/Er\xx&	$-0.85\pm 0.05$	&	$-$0.68	&	7.73	&	$-$0.592	&	11.84	&	2.19 \\
Th/Tm	&	$-0.03\pm 0.04$	&	0.12	&	6.91	&	0.155	&	8.54	&	1.86 \\
Th/Hf	&	$-0.58\pm 0.04$	&	$-$0.20	&	17.50	&	$-$0.036	&	25.16	&	1.86 \\
Th/Os	&	$-1.33\pm 0.16$	&	$-$1.15	&	8.56	&	$-$0.917	&	19.43	&	7.39 \\
Th/Ir	&	$-1.22\pm 0.20$	&	$-$1.18	&	1.87	&	$-$0.839	&	17.78	&	9.50 \\
Th/U	&	$0.90\pm 0.20$	&	0.22	&	14.82	&	0.283	&	13.45	&	4.44 \\
\hline
Th/X (average\xx)	&		&	&	&	&	12.68	&	1.55 \\
\hline
U/Ba\yy &	$-2.24\pm 0.22$	&	\nodata	&	\nodata	&	$-$1.341	&	13.34	&	3.29\\
U/La	&	$-1.38\pm 0.21$	&	$-$0.81	&	8.48	&	$-$0.645	&	10.93	&	3.05\\
U/Ce	&	$-1.79\pm 0.20$	&	$-$1.01	&	11.56	&	$-$1.007	&	11.61	&	3.02\\
U/Pr	&	$-1.25\pm 0.20$	&	$-$0.52	&	10.79	&	$-$0.596	&	9.66	&	2.97\\
U/Nd	&	$-1.95\pm 0.20$	&	$-$1.13	&	12.20	&	$-$1.211	&	10.99	&	2.99\\
U/Sm	&	$-1.61\pm 0.20$	&	$-$0.83	&	11.52	&	$-$1.079	&	7.82	&	3.01\\
U/Eu\yy &	$-1.39\pm 0.20$	&	$-$0.55	&	12.41	&	$-$0.523	&	12.81	&	3.02\\
U/Gd\yy &	$-1.77\pm 0.20$	&	$-$1.03	&	11.04	&	$-$0.852	&	13.68	&	2.99\\
U/Tb	&	$-0.91\pm 0.20$	&	$-$0.33	&	8.64	&	\nodata	&	\nodata	&	3.03\\
U/Dy	&	$-1.81\pm 0.20$	&	$-$1.11	&	10.39	&	$-$1.110	&	10.39	&	3.00\\
U/Ho	&	$-1.11\pm 0.20$	&	\nodata	&	\nodata	&	$-$0.354	&	11.27	&	2.98\\
U/Er\yy &	$-1.75\pm 0.20$	&	$-$0.90	&	12.55	&	$-$0.875	&	12.92	&	3.00\\
U/Tm	&	$-0.93\pm 0.20$	&	$-$0.10	&	12.29	&	$-$0.128	&	11.87	&	2.97\\
U/Hf	&	$-1.48\pm 0.21$	&	$-$0.42	&	15.66	&	$-$0.319	&	17.16	&	3.15\\
U/Os	&	$-2.23\pm 0.25$	&	$-$1.37	&	12.81	&	$-$1.200	&	15.33	&	3.75\\
U/Ir	&	$-2.12\pm 0.28$	&	$-$1.40	&	10.68	&	$-$1.122	&	14.81	&	4.20\\
U/Th	&	$-0.90\pm 0.20$	&	$-$0.22	&	14.82	&	$-$0.283	&	13.45	&	4.44\\
\hline
U/X (average\yy)	&	&		&	&	&	13.19	&	1.53 \\
\hline
Final average\zz	&	&		&	&	&	12.99	&	1.09 \\
\hline
\enddata
\tablenotetext{a}{Abundance ratios used for Th/X average.}
\tablenotetext{b}{Abundance ratios used for U/X average.}
\tablenotetext{c}{Average calculated from ratios marked with $a$, $b$, and U/Th.}

\tablecomments{Initial production ratios (PR: $\eps{X/Y}_0$) are taken from (i)
the \emph{r}-process waiting 
point calculations by \cite{schatz2002} and (ii) the high-entropy wind \emph{r}-process 
models reported by \cite{hill2016}, based on \cite{farouqi2010}.
\explain{small edits made to the Table, to reflect changes in the text.}}
\end{deluxetable}

\listofchanges

\end{document}